\documentclass[journal]{IEEEtran}

\usepackage{balance}

\ifCLASSOPTIONcompsoc
    \usepackage[caption=false, font=normalsize, labelfont=sf, textfont=sf]{subfig}
\else
\usepackage[caption=false, font=footnotesize]{subfig}
\fi

\ifCLASSINFOpdf
   \usepackage[pdftex]{graphicx}
   \graphicspath{{../img/}{../jpeg/}}
   \DeclareGraphicsExtensions{.pdf,.jpeg,.png,.eps}
\else
   \usepackage[dvips]{graphicx}
   \graphicspath{{../img/}{../eps/}}
   \DeclareGraphicsExtensions{.eps}
\fi

\usepackage{tikz}
\usepackage{url}
\usepackage{cite}
\usepackage{color}
\usepackage{float}
\usepackage{xtab,booktabs}

\usepackage[square,sort,comma,numbers]{natbib}
\usepackage[resetlabels,labeled]{multibib}

\newcites{M}{SMS Articles}

\usepackage{xparse}
\let\oriCiteM\citeM

\RenewDocumentCommand{\citeM}{O{} O{} m}{%
  \renewcommand{\citenumfont}[1]{M##1}%
  \oriCiteM[#1][#2]{#3}%
  \renewcommand{\citenumfont}[1]{##1}%
}

\newcommand\copyrighttext{%
  \footnotesize \textbf{Disclaimer}: This work has been accepted for publication in IEEE Transactions of Industrial Informatics. 
  
$\copyright$ 2019 IEEE. Personal use of this material is permitted.  Permission from IEEE must be obtained for all other uses, in any current or future media, including reprinting/republishing this material for advertising or promotional purposes, creating new collective works, for resale or redistribution to servers or lists, or reuse of any copyrighted component of this work in other works.
  }
\newcommand\copyrightnotice{%
\begin{tikzpicture}[remember picture,overlay]
\node[anchor=north,yshift=-2pt] at (current page.north) {\fbox{\parbox{\dimexpr1.15\textwidth-\fboxsep-\fboxrule\relax}{\copyrighttext}}};
\end{tikzpicture}%
}

\newcommand\copyrighttextbottom{%
  \footnotesize Accepted version of B. Rossi and S. Chren, "Smart Grids Data Analysis: A Systematic Mapping Study," in IEEE Transactions on Industrial Informatics. 
  
  DOI: 10.1109/TII.2019.2954098 URL: \url{http://ieeexplore.ieee.org/stamp/stamp.jsp?tp=&arnumber=8903549&isnumber=4389054}
  }
\newcommand\copyrightnoticebottom{%
\begin{tikzpicture}[remember picture,overlay]
\node[anchor=south,yshift=2pt] at (current page.south) {\fbox{\parbox{\dimexpr1.15\textwidth-\fboxsep-\fboxrule\relax}{\copyrighttextbottom}}};
\end{tikzpicture}%
}

\begin{document}
\bstctlcite{MyBSTcontrol} 

\title{Smart Grids Data Analysis: A Systematic\\ Mapping Study}

\author{\IEEEauthorblockN{Bruno Rossi and Stanislav Chren}\\
\IEEEauthorblockA{Faculty of Informatics\\
Masaryk University,
Brno, Czech Republic\\ Email: \{brossi,chren\}@mail.muni.cz} \thanks{The research was supported from European Regional Development Fund Project CERIT Scientific Cloud (No. CZ.02.1.01/0.0/0.0/16\_013/0001802).}
}






\maketitle
\copyrightnotice
\copyrightnoticebottom

\begin{abstract}
Data analytics and data science play a significant role in nowadays society. In the context of Smart Grids (SG), the collection of vast amounts of data has seen the emergence of a plethora of data analysis approaches. In this paper, we conduct a Systematic Mapping Study (SMS) aimed at getting insights about different facets of SG data analysis: application sub-domains (e.g., power load control), aspects covered (e.g., forecasting), used techniques (e.g., clustering), tool-support, research methods (e.g., experiments/simulations), replicability/reproducibility of research. The final goal is to provide a view of the current status of research. Overall, we found that each sub-domain has its peculiarities in terms of techniques, approaches and research methodologies applied. Simulations and experiments play a crucial role in many areas. The replicability of studies is limited concerning the provided implemented algorithms, and to a lower extent due to the usage of private datasets.
\end{abstract}



\section{Introduction}
\IEEEPARstart{T}{he} Smart Grid (SG) is a two-way cyber-physical system utilizing information to provide safe, secure, reliable, resilient, efficient, and sustainable electricity to end-users~\cite{ref:fang2012, ref:chren2016smart}. SGs play nowadays a major role in the integration of the Smart Cities concept by putting into effect the Smart Energy conceptual element: smart electrical energy systems that interconnect utilities and end-users through a Smart Infrastructure~\cite{ref:geisler2013SG-SC,ref:bonetto2017smart,ref:eremia2017smart}. SGs are key enablers, enhancing the decision making process, providing self-healing and automation of the energy grid, and integration of renewable energy sources~\cite{ref:geisler2013SG-SC}.

There are several definitions of a Smart Grid~\cite{ref:clastres2011smart}. The European definition emphasizes the fact that SGs are electricity networks intelligently integrating the behaviour of all actors to reach sustainable, economic and secure energy supply \cite{ref:ESTP2006}. The United States Department of Energy (USDoE) definition focuses more on the security and safety threats to be addressed with resilient and self-healing mechanisms, providing opportunities for new services and markets~\cite{ref:USDoE2004}.

SGs pose several challenges that derive mainly from the integration of the physical infrastructure with information and communication technologies~\cite{ref:cen2012group}. All these challenges need to be addressed with a holistic view taking into consideration all the different layers that form the SG ecosystem~\cite{ref:cen2012group}. Some of the main challenges are the increase of importance for availability of communication networks over traditional confidentiality and integrity aspects in traditional networks~\cite{ref:ancillotti2013role}, the importance of customers' privacy and security of the infrastructure~\cite{ref:mcdaniel2009security}, the relevance of ways to integrate renewable energy sources in a reliable way~\cite{ref:wolsink2012research}, and the usage of information/data available for self-healing and self-monitoring purposes~\cite{ref:amin2008challenges, ref:farhangi2010path,ref:amin2005toward}.

This article is focused on this last challenge, as SGs gave rise to large amount of opportunities in terms of data analytics initiatives: the large availability of data from the smart infrastructure allows many decision support initiatives, but also the implementation of predictive algorithms to improve the provided services \cite{ref:rossi2016anomaly}. Typical examples involve power load forecasting, predicting the  possible load curve that represents the electricity consumed by customers over time \citeM{SMS015:Khan2013b}, or Demand Response (DR) representing load balancing of energy supply and demand during peak hours \citeM{SMS023:Frincu2014}.

The goal of this paper is to provide an overview of data analysis in SGs with a focus on sub-areas, aspects of research, techniques, research methods, tool support, and replicability/reproducibility concerns. We have the following main contributions:
\begin{itemize}
\item a large Systematic Mapping Study (SMS) \cite{ref:petersen2008systematic,ref:barn2017conductingsms} on Smart Grids data analytics, including 359 papers. To our knowledge, this is the largest review in the SG data analysis domain in terms of included articles;
\item a categorization of different SG data analysis sub-domains, with cross-cutting aspects such as techniques used, research methodologies, aspects investigated, replicability concerns with the availability of source code and datasets. While an SMS cannot ensure that all research is covered \cite{ref:budgen2006performing}, it provides a systematic sampling mechanism to look for research aspects holistically;
\end{itemize}

The article is structured as follows. In section II, we provide an overview of Smart Grids concepts relevant for this article: mainly architecture and components that compose the SG infrastructure. In section III, we define the goals, needs, process and research questions for the SMS on data analysis in the SG context. In section IV, we present the results from the overall SMS process, divided into answers to the main research questions. In Section V, we discuss the main threats to validity of the current study. Section VI provides the final conclusions.

\section{Smart Grids}

A SG consists of diverse hardware and software systems with a complex communication infrastructure. To fully understand the smart operations supported by the infrastructure, it is necessary to map the infrastructure to the provided services, usage scenarios, and stakeholders. 

\begin{figure}[!htbp]
\centering
\includegraphics[width=0.95\linewidth]{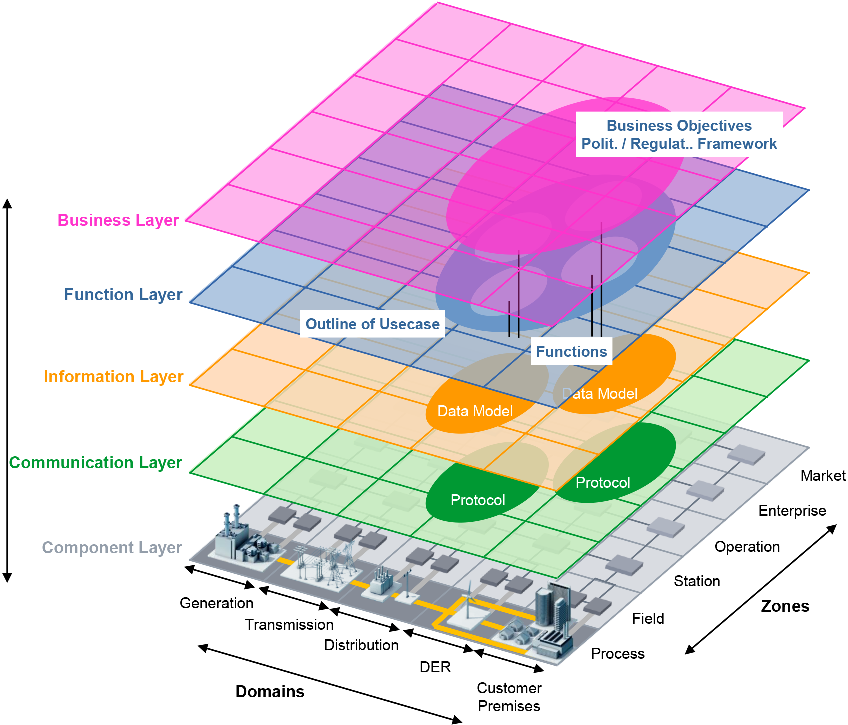}
\caption{The Smart Grid Architecture Model (SGAM) \cite{ref:cen2012group}.}
\label{fig:sgam}
\end{figure}

One of the first attempts to formalize the overall structure of a SG was done by the US National Institute of Standards and Technologies (NIST) which developed a conceptual model of the smart grid \cite{ref:nist2010nist} followed by the more specific reference architecture model \cite{ref:frameworkroadmap}. These models were later modified for the European context by the CEN-CENELEC-ETSI standardization Group \cite{ref:cen2012group}, which resulted in the Smart Grid Architecture Model (SGAM) Framework (Fig.~\ref{fig:sgam}): SGAM is a multi-layered framework that consists of interoperability layers mapped to the SG pane. The SG pane is formed by physical electrical domains and information management zones. The intention of the SGAM model is to represent on which zones of information management the interactions between domains take place. 

The interoperability layers define  viewpoints through which the SG has to be considered \cite{ref:cen2012group}. In the context of our work, the bottom three layers are the most relevant:

The \textit{Component layer} specifies the physical distribution of all participating components in the SG including actors, applications, power system equipment, protection and control devices, network infrastructure, and any kind of devices. The \textit{Communication layer} describes protocols and mechanisms for the exchange of data between components within the context of usage scenario, function or service. The \textit{Information layer} details the information that is exchanged between functions services and components. It contains information objects and the underlying data models.

The \textit{SGAM physical electrical domains} capture the electrical energy conversion chain and consist of 5 parts. Bulk generation represents generation of the energy in large quantities, for example by fossil, nuclear or hydro-power plants. Such generation is connected to the transmission system. Transmission represents the infrastructure and organization for long-distance energy transportation. Distribution represents the infrastructure and organization that distributes the electricity to customers. Distributed Energy Resources (DER) describes the distributed electrical resources connected to the public distribution grid. Customer premises include the consumers of electricity and the local producers.

\begin{figure}[!htbp]
\centering
\includegraphics[width=0.6\linewidth]{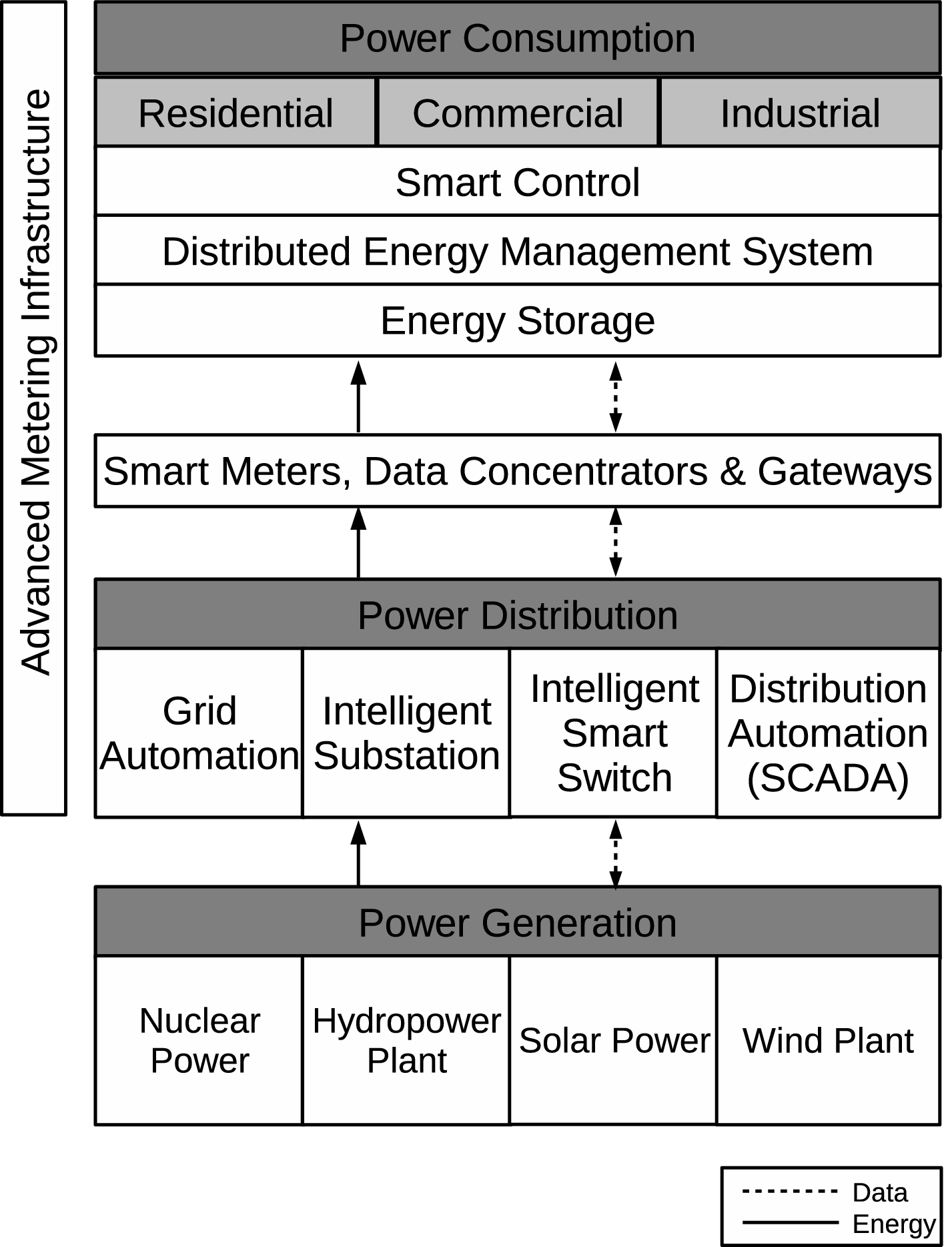}
\caption{Smart Grid data and information flow. Adapted from \cite{ref:daki2017big}.}
\label{fig:sg-diagram}
\end{figure}

An important feature of SGs is better integration of renewable energy sources into the system and more refined management of energy consumption and production. This is achieved by the bidirectional flow of both energy and data between power generation, distribution and consumption (Fig.~\ref{fig:sg-diagram}). Power generation is possible through various sources such as nuclear, hydropower and renewable. Communication with power distribution connects consumers with the electricity grid and transmits data using the Advanced Metering Infrastructure (AMI). Technologies for autonomous infrastructure monitoring and supervision are very important in this context. Power consumption involves the final users of electricity, both residential and industrial.

From the power distribution and the customer perspective, the main services of the SG are provided by the Advanced Metering Infrastructure (AMI) system. It allows collecting measurements about energy consumption and production, which help utilities control the energy and be cost and time efficient. AMI enables dynamic load management and various demand-response programs which involve dynamic pricing of energy and remote appliance control. The core components of AMI are smart meters, data concentrators, gateways, and head-end systems. 

The smart meter is a replacement for the traditional electricity meters installed at customer premises. A smart meter is expected to have capabilities of real-time or near real-time capture of electricity usage, remote reading and controllability, linking to other commodity supply (gas and water), ability to capture events (e.g., power quality), being interoperable within an SG environment (e.g. as specified by NIST~\cite{ref:nist2010nist} and the SGAM framework~\cite{ref:cen2012group})~\cite{ref:van2006smart}. The smart meters can exchange data with the power distribution company in several ways: directly through a mobile operator network, through a data concentrator, or through a gateway~\cite{ref:chren2016smart}. 

A number of various technologies are involved in the communication within the AMI. The used technologies need to take into account communication requirements of various usage scenarios, such as maximum allowed latency, payload size, and frequency of data transfer. 

Reliability plays a key role in the smooth operation of the SG infrastructure. Reliability is ensured by different components/systems \cite{ref:chren2018SCSP}, such as the Blackout Prevention System (WAMPAC) that utilizes Phasor Measurement Units (PMUs) to obtain relevant information from the grid. PMU is a device deployed in the transmission network which measures electrical waves in a grid and helps to detect anomalies and failures. Furthermore, the Supervisory Control and Data Acquisition System (SCADA) is one of the core systems that provides support to operation activities and functions in transmission automation, dispatch centers and control rooms.

\section{SMS}
To survey SG data analysis research, we followed the Systematic Mapping Study Methodology (SMS)~\cite{ref:petersen2008systematic,ref:barn2017conductingsms}. An SMS is useful to explore a research area by identifying the amount and frequency of publications over time to see trends, the type of research and main results available. It provides a visual summary that can be considered a map of existing research~\cite{ref:petersen2008systematic,ref:barn2017conductingsms}.
There are differences between an SMS and another popular formal review methodology called Systematic Literature Review (SLR)~\cite{ref:kitchenham2004procedures,ref:petersen2008systematic,ref:petersen2015guidelines}: an SLR focuses more on the quality and findings of research, thus providing a smaller set of research results. An SMS allows researchers to consider a larger number of articles as they are not evaluated under such level of detail. In summary, an SMS is providing a more coarse-grained overview, at the expense of the in-depth analysis provided by an SLR~\cite{ref:petersen2008systematic}. For the goals of this review, we considered the SMS methodology more appropriate to get a wider view of the research area.

\begin{figure*}[!htbp]
\centering
\includegraphics[width=0.7\linewidth]{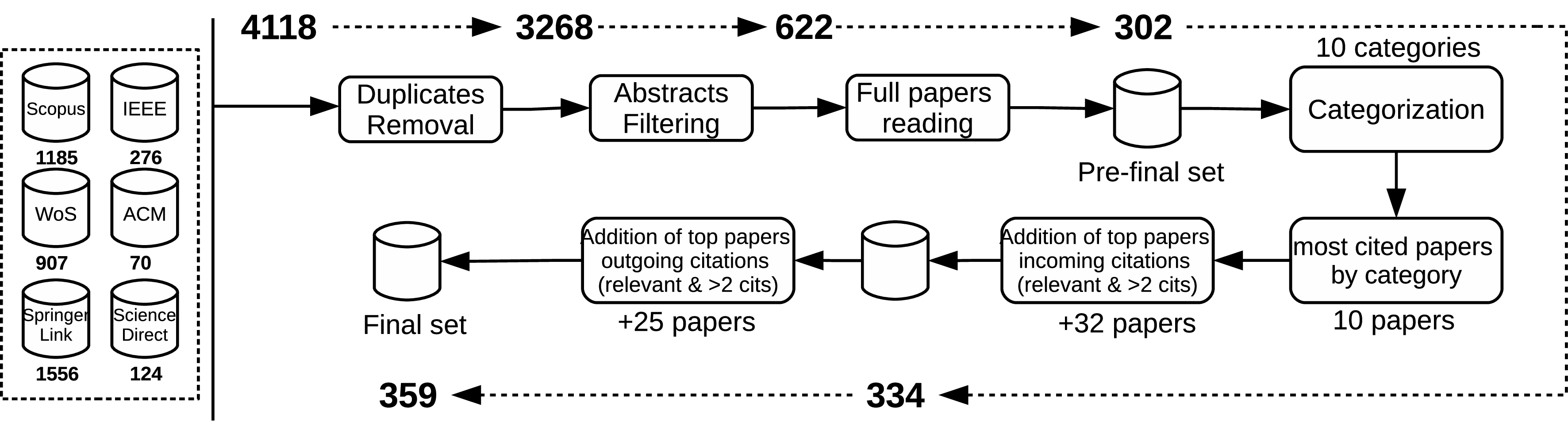}
\caption{The Systematic Mapping Study filtering process with number of articles at each stage.}
\label{fig:sms-process}
\end{figure*}

\subsection{SMS Needs}
\label{sec:SMS-needs}
There are several existing reviews about data analytics in Smart Grids (\cite{ref:diamantoulakis2015big, ref:tu2017bigdatareview, ref:hu2016bigdatachallenges, ref:alahakoon2016futuresurvey,ref:benetti2016slr, ref:yan2013communicationsurvey, ref:bhardwaj2017smartgridsanalyticssystematic, ref:pawar2017data-analytics-brief-review, ref:ge2018big}). However, compared to the goals of the current article, they mostly focused on the Big Data level (e.g., \cite{ref:diamantoulakis2015big, ref:tu2017bigdatareview}) on a specific sub-domain, like data analysis for smart meters \cite{ref:alahakoon2016futuresurvey}, demand-load response \cite{ref:benetti2016slr}, or based on limited set of resources \cite{ref:bhardwaj2017smartgridsanalyticssystematic, ref:pawar2017data-analytics-brief-review}. 
Furthermore, we reviewed existing systematic literature reviews and mapping studies in the area of data analytics for the Smart Grids, but we could not find comparable prior studies, existing ones being either focused on Smart Cities \cite{ref:cocchia2014smartcity}, or the aforementioned specific aspects (e.g., demand-load response \cite{ref:benetti2016slr}).
The article that has more similarities with the current one is the valuable review from Alahakoon \textit{et al.} \cite{ref:alahakoon2016futuresurvey}, covering Smart Meter data analysis. However, while the review of Alahakoon \textit{et al.} \cite{ref:alahakoon2016futuresurvey} is focused on smart meters and data analysis to provide business intelligence in different aspects (e.g., pricing), our review complements it by considering a broader context (the whole SG layers). Furthermore, by applying the SMS process we include a much larger set of references that are discussed in aggregated form to provide numerical evidence about the most researched aspects and utilized techniques.

\begin{table*}[!t]
\renewcommand{\arraystretch}{1.3}
\caption{Queries Run on Different Repositories (mined on 20/08/2019)}
\label{tbl:queries}
\scriptsize
\centering
\begin{tabular}{|p{2.8cm}||p{12cm}|p{1.5cm}|}
\hline
Repository & Query & \# Found  \\
\hline
IEEExplore & 
"Abstract":"data" AND "Abstract":"analysis" AND ("Abstract":"smart grid" OR "Abstract":"smart grids") AND ("Abstract":"algorithm" OR "Abstract":"method") & 276  \\
\hline  
ACM & 
"query": \{ data AND analysis AND ("smart grid" OR "smart grids") AND (algorithm OR method) \}  & 70 \\ 
\hline
ScienceDirect  & 
TITLE-ABSTR-KEY(data AND analysis AND ("smart grid" OR "smart grids") AND (algorithm OR method)
 & 124 \\ 
\hline
Scopus  & 
 TITLE-ABS-KEY ( data  AND  analysis  AND  ( algorithm  OR  method )  AND  "smart grid*" )  AND  PUBYEAR  $>$  2007  AND  ( LIMIT-TO ( LANGUAGE ,  "English" ) )  AND  ( LIMIT-TO ( DOCTYPE ,  "cp" )  OR  LIMIT-TO ( DOCTYPE ,  "ar" )  OR  LIMIT-TO ( DOCTYPE ,  "ch" ) ) 
 & 1185 \\ 
\hline
Web of Science (WoS)  & 
(TS=(data AND analysis AND (algorithm OR method) AND smart grid*) AND PY=(2007-2019)) AND LANGUAGE: (English) AND DOCUMENT TYPES: (Article OR Book Chapter OR Proceedings Paper)
 & 907 \\ 
\hline
SpringerLink  & 
data AND analysis AND ("smart grid" OR "smart grids") AND (algorithm OR method) AND (dataset OR "data set") ENGLISH (2007-2019) & 1556\\ 
\hline
\end{tabular}
\end{table*}

\subsection{SMS Process}
To conduct the SMS, we followed guidelines in \cite{ref:petersen2008systematic,ref:barn2017conductingsms}.
The goal of the SMS was to get an overview of the area of data analysis in the context of Smart Grids. This is a large  area with multiple aspects considered, with limited studies that provide a bottom-up review (section \ref{sec:SMS-needs} SMS Needs). We followed the process represented in Fig. \ref{fig:sms-process}. Each phase has indication of the number of articles that were included. After the definition of the research questions, we defined set of queries that were run on a set of six digital repositories (yielding 4118 articles), we removed then duplicates (3268 articles remaining), and identified relevant articles based on inclusion/exclusion criteria applied to abstracts (622 remaining), that were further filtered by full papers reading, reaching a pre-final set (302 articles remaining). Categories of articles were extracted by means of open coding, and outgoing/incoming citation for the most cited articles in each category were used to improve the quality of the results (+57 articles added). The final set of 359 articles was used to summarize the overall research area. To drive the SMS process, we set the following research questions:

\begin{enumerate}
\item[{RQ}1.] \textit{What are \textbf{common aspects} that are discussed in the identified sub-domains?}
\item[{RQ}2.] \textit{Which are the reported \textbf{most used techniques} in the identified sub-domains?}
\item[{RQ}3.] \textit{Which are the most used \textbf{software tools / development environments} used for data analysis in the identified sub-domains?}
\item[{RQ}4.] \textit{What is the status of replicability / reproducibility of the studies in terms of \textbf{datasets used} and \textbf{availability of implemented algorithms}?}
\end{enumerate}

\subsection{SMS Queries}
A relevant part of the search process execution is the selection of the digital repositories. We selected six repositories that we consider relevant for the type of search.
\begin{enumerate}
\item[{DR}1.] IEEExplore (\url{http://ieeexplore.ieee.org})
\item[{DR}2.] ACM Digital library (\url{https://dl.acm.org})
\item[{DR}3.] Elsevier ScienceDirect (\url{www.sciencedirect.com}) 
\item[{DR}4.] Scopus (\url{https://www.scopus.com})
\item[{DR}5.] Web of Science (WoS) (\url{www.webofknowledge.com})
\item[{DR}6.] SpringerLink (\url{https://link.springer.com})
\end{enumerate}

For each repository we run the queries either on abstracts or combination of abstracts, keywords and titles: only exception is the SpringerLink repository that allowed only a full-text search (Table \ref{tbl:queries}).

\subsection{Inclusion / Exclusion Criteria}

Inclusion criteria were applied to each reviewed article, looking at the abstract, and if necessary to the full-text:

\begin{enumerate}
\item[{IC}1.] Empirical studies that included either the application to real cases/datasets or simulation/numerical experiments. No pure theoretical papers were included;
\item[{IC}2.] Studies about the application of techniques either to case studies, experiments, or simulations;
\item[{IC}3.] Articles newer equal than year 2007 (included);
\item[{IC}4.] Only conference/workshops/journal papers included;
\item[{IC}5.] English language only;
\end{enumerate}

\noindent Exclusion criteria were applied based on the type of article and formats, but also content-wise:

\begin{enumerate}
\item[{EC}1.] Papers not dealing with the SG domain;
\item[{EC}2.] No papers about electric vehicles data analysis;
\item[{EC}3.] No smart cities related papers;
\item[{EC}4.] No wireless sensor network papers;
\item[{EC}5.] No formats such as presentations, slides, posters;
\item[{EC}6.] No grey literature (e.g., editorials, reports, keynotes) 
\end{enumerate}

\section{SMS Results}

\subsection{Category Extraction}
\label{sec:category-ext}
We followed a process of open coding by reading the abstracts from all the articles selected and mapping them to categories. The process was to create categories and merge/expand them, depending on new articles that were analyzed. While some papers could fit in multiple categories, we categorized them to the category that was more representative of the main focus of the article (see section \ref{sec:category-ext}).

For the categorization, we took also into consideration previous categorizations of articles, in particular by Alahakoon \textit{et al.} \cite{ref:alahakoon2016futuresurvey} that was focused on Smart Meters intelligence, which considered "customer profiling", "segmentation", "cluster analysis", "load forecasting", "pricing intelligence", "capturing irregularities", and "metering intelligence to support real-time operations" as main metering intelligence activities. In our case, \textit{C3. events analysis} can be roughly considered as the category of supporting real-time intelligence in Alahakoon \textit{et al.} \cite{ref:alahakoon2016futuresurvey}. The category capturing irregularities \cite{ref:alahakoon2016futuresurvey} can be seen as both our categories \textit{C9. security} and \textit{C10. SG failures}, as it contains techniques related to the identification and detection of anomalies that could lead to failures. We preferred to have two distinguished categories, one dealing with identification of security threats, the other about availability and reliability of the SG infrastructure to give a better distinction about the main research focus. We added \textit{C8. privacy}, as many articles were focused on privacy issues and schemes for data anonymization while retaining relevant information.

At the end of the open coding process, we created ten macro-categories derived from the papers: 

\begin{enumerate}
\item[C1.] \textbf{Customer Profiling}: Classification/clustering of users in common classes according to common characteristics (e.g., usage
of appliances);
\item[C2.] \textbf{Energy output forecasts}: Prediction of energy output from renewable energy resources (variable in time);
\item[C3.] \textbf{Events analysis}: Analysis of logs/events generated at different levels of the Smart Grids infrastructure (e.g., to detect anomalies);
\item[C4.] \textbf{Load segregation}: Disaggregating information about energy consumption on an appliance-by-appliance basis;
\item[C5.] \textbf{Power loads / consumption analysis}: Predicting the power consumption with the ultimate goal of reaching  balance of supply and demand in the power market;
\item[C6.] \textbf{Power quality}: Power disturbance classification and algorithms for countermeasures and data compression;
\item[C7.] \textbf{Pricing}: Dynamics of forecasting electricity price and demand;
\item[C8.] \textbf{Privacy}: Data anonymization algorithms and other concerns related to disclosing private information about consumers;
\item[C9.] \textbf{Security}: Algorithms dealing with countermeasures/prevention of attacks to the SG infrastructure;
\item[C10.] \textbf{Smart Grid Failures}: Aspects of SG failures, faults, and countermeasures;
\end{enumerate}

One category requires additional clarifications: \textit{C3.~events analysis}: we clustered in this category all the articles which were dealing with algorithms for stream/event processing inside the SG infrastructure with no specific goal identified in the other categories.

\begin{table*}[!htbp]
\renewcommand{\arraystretch}{1.3}
\caption{Main aspects discussed ordered by number of occurrences}
\label{tbl:aspects-mapping}
\centering
\begin{tabular}{|p{2.5cm}||p{14.5cm}|}
\hline
\textbf{Category} & \textbf{Aspects} \\
\hline
C1. Customer Profiling & 1.~Power consumption profile clustering (\citeM{SMS004:Flath2012,SMS009:Liu2012,SMS044:Wang2015c,SMS049:Benitez2016,SMS059:Hopf2016,SMS068:Lu2016a,SMS069:Peng2016758,SMS077:Abdulaal20171050,SMS082:Jiang2017658,SMS092:Tornai201725,SMS093:Tornai2017,SMS124:Fabisz2014,SMS170:Li2016b,SMS182:Viegas2016,SMS221:rhodes2014clustering,SMS246:rasanen2010data,SMS147:Granell20153217,SMS027:Abdulaal2015133,SMS031:Buitrago2015167,019:Chen201849,034:Kojury-Naftchali20181,060:Barbour:2018:EHL:3276774.3276793,064:Chou2018709}), 2.~power consumption pattern recognition (\citeM{SMS007:Labeeuw2012a,SMS024:Ming2014, SMS028:Alhamoud2015, SMS033:Chen2015e,SMS076:Zhu2016704,SMS083:Justo2017,SMS099:Zhou201773,SMS123:Dent2014,SMS227:lavin2015clustering,SMS232:do2016cluster,SMS032:Chelmis2015,SMS055:Cugliari2016,SMS073:Xu2016,028:ISI:000457952500041,079:Butunoi20181}), 3. events/tasks extraction (\citeM{SMS001:Cho20102011}) \\
\hline
C2. Energy output forecast & 
1.~forecast renewable power sources (\citeM{SMS043:Wang2015b,SMS109:Hosoda2012119,SMS119:Kramer2013,SMS133:Pravilovic2014,SMS138:Bessa2015232,SMS142:DeLeone2015,SMS150:Heinermann2015,SMS158:Zhang2015a,SMS162:Chen2016b,SMS176:Oprea2016,SMS194:GonzalezOrdiano2017,SMS222:bessa2015probabilistic,SMS225:filipe2015hybrid,SMS235:dowell2016very,SMS239:zhang2016spatial,SMS252:fernandez2012short,SMS255:pedro2012assessment,SMS257:yang2012novel,SMS261:huang2013forecasting,SMS062:Jiang2016b,031:abdel-nasser_accurate_2019,083:ceci_spatial_2019,092:chen_wind_2018}) \\ 
\hline
C3. Events analysis & 
1.~Data stream processing (\citeM{SMS020:Chen2014a,SMS106:Guarracino2012,SMS174:Maurya2016a,SMS223:dahal2015event}), 2.~clustering events (\citeM{SMS084:Klinginsmith201775,SMS167:Klinginsmith2016233}), 3.~anomaly detection (\citeM{SMS155:Marnerides2015,SMS169:Lazzaretti2016a,SMS126:Fu2014,SMS135:Weng201467,006:Nur:2019:CAC:3307772.3331028,049:sial_detecting_2019,051:ISI:000457998600001,066:chong_heuristics-based_2018,077:luo_real-time_2018}), 4.~recommendation for energy utilization (\citeM{SMS060:Hosoe2016}), 5.~smart grid partitioning (\citeM{SMS117:Ilic2013a,052:blasch_dynamic_2018,052:blasch_dynamic_2018}) \\
\hline
C4. Load segregation & 
1.~Non-intrusive appliance load monitoring ( \citeM{SMS110:Kramer2012,SMS134:Semwal2014,SMS140:Bonfigli20151175,SMS144:Dinesh2015,SMS146:Gonzalez-Pardo2015,SMS166:Jimenez2016890,SMS178:Pijnenburg20163408,SMS122:Ziekow2013,SMS168:Koutitas20161665,SMS212:zeifman2012disaggregation,SMS249:kim2011unsupervised,030:yang_systematic_2019,038:Wang2019,072:Laurinec2018210})\\
\hline
C5. Power loads / consumption analysis & 
1.~power loads forecasting (\citeM{SMS008:Lee2012b,SMS021:Fan20141,SMS023:Frincu2014,SMS026:Yang2014,SMS030:Beaude2015,SMS052:Campos2016,SMS061:Islam2016,SMS086:Lau2017,SMS094:Usha2017,SMS098:Yu2017738,SMS141:Dagnely2015,SMS159:Ali2016,SMS186:Alkhatib2017,SMS187:Alkhatib2017a,SMS202:Moon2017,SMS233:chou2016time,SMS241:bassamzadeh2017multiscale,SMS242:fan2017short,SMS245:li2010classification,SMS002:Hong2010212,SMS005:Haider2012,SMS014:Hernandez2013,SMS015:Khan2013b,SMS019:Aparicio2014a,SMS025:Semeraro2014,SMS037:Niska2015,SMS041:Son2015,SMS058:Hassan2016,SMS066:Liu2016k,SMS070:Rana2016118,SMS087:Li2017a,SMS088:Lu2016b,SMS016:Niska2013,SMS114:Gadd2013,SMS153:Idowu2015554,SMS164:Dufour2016a,SMS101:Kramer2010,SMS107:Hassan2012,SMS116:Hou2013170,SMS118:Jain2013a,SMS120:Marinescu201325,SMS125:Fattaheian-Dehkordi20141650,SMS130:Loewenstern2014,SMS139:Bianchi20151931,SMS145:Dreiseitl2015,SMS148:Hoeverstad2015,SMS149:Hayes2015a,SMS151:Hsiao201533,SMS156:Pellegrini2015,SMS160:Chemetova2016340,SMS163:Dhillon2016a,SMS171:Li2016d,SMS180:Sreekumar2015,SMS185:Alberg2017,SMS193:Gerwig2017,SMS195:He2017254,SMS197:Lei2017,SMS198:Li2017,SMS199:Lis2017,SMS200:Liu2017730,SMS204:Rego2017,SMS206:Sun2017,SMS207:Vrablecova2017,SMS208:Wan2017,SMS229:shamshirband2015heat,SMS237:magyar2016risk,SMS211:llanos2012load,SMS240:zhou2016residential,SMS259:borges2013evaluating,001:nguyen_deepenergy:_2018,005:sun_short-term_2018,007:Dong20181,008:MOHAN2018229,009:rabie_fog_2019,014:kaur_hybrid_2019,029:ISI:000445532600001,035:Mohammad20181771,043:ISI:000426734600194,047:Wang20196446,048:laurinec_density-based_2019,055:HE2019565,056:ISI:000460746500007,059:gan_embedding_2018,067:Song:2018:HDL:3208903.3210283,068:Laurinec2019413,069:li_long-term_2018,075:sun_power_2019,080:ke_short-term_2019,081:dash_short-term_2019,082:ISI:000425074300010,086:gao_probabilistic_2018-1}), 2.~power loads clustering (\citeM{SMS006:Hernandez2012,SMS017:Pereira2013,SMS039:Pal2015,SMS075:Zhang2016f,SMS090:Paisios2017,SMS203:Natividad2017,SMS205:Sharma2017,SMS003:Kim20111102,SMS103:Kim2011,SMS111:Ramos2012a,SMS175:Mets2016,SMS210:Zehetbauer2017,011:ISI:000431531400006,016:ISI:000468028302036,024:Gao2019253,036:Lin2018,044:ISI:000454335100038,045:Fang2018455,046:Fu201876,058:8652444,065:Yang2018,087:Manojlovic2019140}), 3.~consumption data analysis and modelling (\citeM{SMS022:Faria2014a,SMS157:Yu2015a,SMS190:Bharathi2017,SMS214:cho2013electric,SMS226:hsu2015identifying,SMS231:benedetti2016energy,SMS053:Cataliotti2016,SMS132:Panapakidis2014137,SMS079:Balakrishna2017,SMS188:Andersson2017}) \\
\hline
C6. Power quality & 
1.~power quality disturbances classification (\citeM{SMS011:Barbosa2013,SMS050:Borges2016824,SMS243:ericsti2010wavelet,SMS248:decanini2011detection,SMS250:lee2011optimal,SMS253:huang2012power,SMS256:rodriguez2012rule,SMS258:biswal2013measurement,SMS260:he2013real,SMS264:biswal2014automatic,SMS067:Liu2016e,SMS131:Majidpour2014,SMS177:Peppanen2016,SMS051:Botev2016,020:ISI:000458942800091}), 2.~power data compression (\citeM{SMS035:Eichinger2015a,SMS036:Khan2015a,SMS102:Das201195,SMS128:Khan2014a,SMS183:Wang2016,SMS209:Wang20172142,SMS216:mehra2013modes,SMS218:top2013compressing,SMS234:cormane2016spectral,SMS244:klump2010lossless,SMS251:ning2011wavelet,032:de_andrade_advances_2019,040:shamachurn_assessing_2019,073:parvez_online_2018,090:Bhuiyan2018}), 3.~measurement errors (\citeM{010:ISI:000445054400035,037:ISI:000454332000012,071:huang_missing_2019}), 4.~meter placement for quality estimation (\citeM{SMS046:Ali20161552,SMS112:Abdel-Majeed2013})\\ 
\hline
C7. Pricing & 1.~pricing forecasting (\citeM{SMS010:Mori2012c,SMS071:Shrivastava2016,SMS072:Wang2016i,SMS136:Alamaniotis2015,SMS238:monteiro2016short,SMS247:tan2010day,SMS254:motamedi2012electricity,SMS262:shayeghi2013day,SMS263:wu2013new,SMS265:shayeghi2015simultaneous,SMS266:panapakidis2016day,SMS267:shayeghi2017day,SMS056:Ghasemi201640,SMS108:He2012230,013:rogers_genetic_2019,015:luo_hybrid_2018,018:ghadimi_new_2018,033:Azdemir2018224,041:barolli_big_2019-1,054:Almahmoud20193415,061:MUJEEB2019101642,078:ISI:000427121200015,089:ISI:000426287900006}), 2.~pricing impact on customer behaviour (\citeM{SMS042:Waczowicz2015a,SMS137:Berlink2015,017:Yang20193374,026:Li2019117,039:ISI:000447404600007,074:Saez-Gallego20185005,076:REPL-Li2018-Principalcomponentanalysis}) \\
\hline
C8. Privacy & 1.~privacy preserving data aggregation (\citeM{SMS013:He201367,SMS040:Savi20152409,SMS080:Bao2017,SMS081:Guan2017a,SMS095:Vahedi201728,SMS184:Afrin2017,SMS196:Knirsch2017a,SMS219:tudor2013analysis,SMS236:gulisano2016bes,SMS115:Ge2013,SMS192:Engel20171710,SMS215:ford2013clustering,SMS179:Salinas2016,SMS181:Unterweger2016,027:ISI:000451814000110,063:Ge2018966}), 2.~data re-identification (\citeM{SMS113:Buchmann2013,SMS230:tudor2015study}) \\
\hline
C9. Security & 1.~false data injection attacks (\citeM{SMS045:Yu20151219,SMS047:An2016240,SMS048:Anwar2016180,SMS065:Landford2016,SMS089:Mohammadpourfard2017242,SMS091:Tang2017172,SMS096:Xu2017a,SMS127:Hao2014,SMS129:Li2014a,SMS152:Hu2015,SMS189:Anwar201758,SMS191:Bhattacharjee2017,SMS201:Mishra2017,SMS228:liu2015collaborative,SMS012:Esmalifalak2013808,021:Chen201973,022:Zhong2018,050:8646454,053:8791598,070:ISI:000449541500001}),
2.~intrusion detection (\citeM{SMS018:Raciti2013,SMS063:Kosek2016,SMS078:Andrysiak2017,SMS104:Choi2012,SMS105:Faisal2012,SMS154:Krishna2015,SMS213:ali2013configuration,SMS224:faisal2015data,SMS029:Anwar2015a}),  3.~energy theft (\citeM{SMS054:Cody20161175,SMS057:Han2016a,SMS074:Zanetti2016,SMS097:Yip2017230,SMS121:Nikovski2013,SMS172:Liu2016h,003:ganguly_novel_2018,025:Razavi2019481,042:ISI:000443697900010,057:ISI:000424131500014,084:8422731,085:Jiao2018,091:ISI:000429266400030}) \\
\hline
C10. SG failures & 
1.~fault status detection \citeM{SMS034:DeSantis2015b,SMS085:Kordestani2017,SMS100:Cai2010642,SMS143:DeSantis2015a,SMS165:Jiang20162525,SMS173:Mahfouz2016,SMS220:jiang2014fault,SMS161:Chen20161726,012:Ziga2019507,023:ISI:000425203500024,062:deSouzaPereira2018640,088:Jana2018387}, 2.~power distribution reliability \citeM{SMS064:Kuhi2016,SMS217:nunez2013feature,002:sun_distributed_2018,004:zitouni_predictive_2019}, 3.~fault type classification \citeM{SMS038:Oubrahim20152735}
\\
\hline
\end{tabular}
\end{table*}

\subsection{RQ1. What are common aspects that are discussed in the identified sub-domains?}
The common aspects discussed in the articles can be useful to characterize the different research areas, by providing general themes. The main aspects discussed in each paper are presented in Table \ref{tbl:aspects-mapping}. 

For \textit{C1. Customer profiling}, the category is focused on determining customer profiles based on power load consumption. For example, load profile clustering is the focus of Flath \textit{et al.} \citeM{SMS004:Flath2012}, in which smart meter data is clustered by means of the k-means algorithm to identify group of customers to provide more targeted services. Power consumption pattern recognition is the focus of Ming \textit{et al.} \citeM{SMS024:Ming2014}, by extracting common behaviour of users from power consumption traces and clustering them by means of an enhanced k-means approach. Using customers profiles for power load forecasting by means of wavelet transforms and various clustering algorithms was provided in  Cugliari \textit{et al.} \citeM{SMS055:Cugliari2016}. Extracting fine-grained customers activity profiles was researched in Cho \textit{et al.} \citeM{SMS001:Cho20102011}.

The C1 category can be considered a parallel category to \textit{C5. Power load/consumption analysis}, but category \textit{C1} is more focused on clustering and consumption patterns identification, while in \textit{C5} the focus is more on classification and prediction of power consumption loads. \textit{"Smart Meter data"} plays a major role both in \textit{C1} and \textit{C2} categories, being the main source of data for analysis. For both categories, energy/power consumption related concepts are the main focus.

For \textit{C2. Energy output forecasts}, forecasting renewable power sources and power indicator forecasts are the most common themes. The category is more focused on (wind/solar) power/energy generation forecasting, showing relevant concepts in techniques such as moving averages, and regression analysis. For example, Wang \textit{et al.}~\citeM{SMS043:Wang2015b} provide models for forecasting renewable power generation sources, based on local linear models for nonlinear timeseries. Jiang \textit{et al.}~\citeM{SMS062:Jiang2016b} provide a model that  combines denoising methods and optimization algorithms to forecast power outputs.

\textit{C3. Events analysis} collects more variety of articles, covering data stream processing (like Chen \textit{et al.}~\citeM{SMS020:Chen2014a} that provide big data classification for data streams in SG), clustering of events (like Klinginsmith \textit{et al.}~\citeM{SMS084:Klinginsmith201775} that propose a comparison of different clustering algorithms for SG events), anomaly detection (like Fu \textit{et al.} \citeM{SMS126:Fu2014} identifying efficiently critical events from smart meters for online learning), recommendations for energy utilization (like Hosoe \textit{et al.}~\citeM{SMS060:Hosoe2016}, suggesting a recommendation system based on events of home appliances usage), smart meters grouping (like Ilic \textit{et al.}~\citeM{SMS117:Ilic2013a} grouping smart meter events for forecasting). The category is more focused on \textit{"data streams"} and \textit{"critical events"} with sources deriving from PMU data. Clustering techniques are often relevant for the data analysis process in this category. 

\textit{C4. Load segregation} is focused on Non-intrusive Appliance Load Monitoring (NIALM). Examples are Kramer \textit{et al.}~\citeM{SMS110:Kramer2012} that propose an ensemble of models for appliances identification based on power consumption, and Koutitas \textit{et al.}~\citeM{SMS168:Koutitas20161665} that decompose smart meter data to associate the data points with the operation of the smart home appliances by using fuzzy logic and pattern recognition. 

\textit{C5. Power loads / consumption} is focused on consumption clustering, prediction, consumption data analysis and modelling.
For example, Lee \textit{et al.}~\citeM{SMS008:Lee2012b} use a neural network model for a power consumption prediction model for demand forecasting. Hernandez \textit{et al.}~\citeM{SMS006:Hernandez2012} cluster energy consumption patterns in industrial parks using Self Organizing Maps and k-means clustering, while Faria \textit{et al.}~\citeM{SMS022:Faria2014a} determine the expected consumption per customer, linking the consumption to the commercial losses.

\begin{table*}[!htbp]
\renewcommand{\arraystretch}{1.3}
\caption{Main techniques applied ordered by number of occurrences (one paper can apply more techniques)}
\label{tbl:techniques-mapping}
\scriptsize
\tabcolsep=0.11cm
\begin{tabular}{|p{1.5cm}|p{7.0cm}p{8.5cm}|}
\hline
\textbf{Category} & \textbf{Main Techniques} & \textbf{Other Techniques} \\
\hline
C1. Customer Profiling& 


\textbf{k-means clustering (19)} (\citeM{SMS004:Flath2012,SMS044:Wang2015c,SMS069:Peng2016758,SMS077:Abdulaal20171050,SMS082:Jiang2017658,SMS170:Li2016b,SMS221:rhodes2014clustering,SMS232:do2016cluster,SMS246:rasanen2010data,SMS147:Granell20153217,SMS027:Abdulaal2015133,SMS031:Buitrago2015167,SMS033:Chen2015e,SMS123:Dent2014,SMS227:lavin2015clustering,SMS032:Chelmis2015,028:ISI:000457952500041,034:Kojury-Naftchali20181,060:Barbour:2018:EHL:3276774.3276793}), \textbf{fuzzy c-means clustering (7)} ( \citeM{SMS044:Wang2015c,SMS068:Lu2016a,SMS069:Peng2016758,SMS033:Chen2015e,SMS076:Zhu2016704,SMS099:Zhou201773,SMS123:Dent2014}), \textbf{Hierarchical Clustering (HAC) (7)} (\citeM{SMS069:Peng2016758,SMS092:Tornai201725,SMS246:rasanen2010data,SMS147:Granell20153217,SMS123:Dent2014,SMS032:Chelmis2015,SMS055:Cugliari2016}), \textbf{Support Vector Machine (SVM) (7)} \citeM{SMS059:Hopf2016,SMS069:Peng2016758,SMS077:Abdulaal20171050,SMS182:Viegas2016, SMS007:Labeeuw2012a,SMS073:Xu2016,064:Chou2018709} 
&
\tiny{
Self-Organising Map (SOM) (5) \citeM{SMS009:Liu2012,SMS246:rasanen2010data,SMS027:Abdulaal2015133,SMS123:Dent2014,034:Kojury-Naftchali20181},
Multi Layer Perceptron (MLP) (4) \citeM{SMS007:Labeeuw2012a,SMS092:Tornai201725,SMS031:Buitrago2015167,064:Chou2018709}, k-Nearest Neighbour (kNN) \citeM{SMS059:Hopf2016,SMS069:Peng2016758,079:Butunoi20181}, DBSCAN clustering \citeM{079:Butunoi20181}, Random Forest \citeM{SMS069:Peng2016758,SMS028:Alhamoud2015,SMS123:Dent2014}, Principal Component Analysis (PCA) \citeM{SMS068:Lu2016a,SMS032:Chelmis2015,028:ISI:000457952500041,034:Kojury-Naftchali20181}, Decision Trees \citeM{SMS077:Abdulaal20171050,SMS007:Labeeuw2012a}, Linear regression \citeM{060:Barbour:2018:EHL:3276774.3276793,064:Chou2018709}, Logistic Regression \citeM{SMS232:do2016cluster,SMS232:do2016cluster}, multi-resolution clustering (MRC) \citeM{SMS170:Li2016b}, Wavelet-based clustering \citeM{SMS055:Cugliari2016}, ART Neural Network \citeM{SMS083:Justo2017}, t-means clustering \citeM{SMS024:Ming2014}, dynamic clustering with Haussdorff similarity distance \citeM{SMS049:Benitez2016}, Ensemble Methods \citeM{SMS077:Abdulaal20171050,019:Chen201849}, Genetic Algorithms \citeM{SMS077:Abdulaal20171050}, Discriminant Analysis \citeM{SMS077:Abdulaal20171050}, Discrete Wavelet Transform \citeM{SMS082:Jiang2017658,019:Chen201849}, Deep Learning CNNs \citeM{SMS093:Tornai2017}, statistical-based \citeM{SMS124:Fabisz2014}, Takagi-Sugeno fuzzy models (FM) \citeM{SMS182:Viegas2016}, binary regression analysis \citeM{SMS221:rhodes2014clustering}, Dirichlet Process Mixture Model (DPMM) \citeM{SMS147:Granell20153217}, Apriori algorithm \citeM{SMS028:Alhamoud2015}, Dynamic Time Warping \citeM{028:ISI:000457952500041} }
\\
\hline
C2. Energy output forecast & 

\textbf{Autoregressive Forecasting Model (7)} (\citeM{SMS138:Bessa2015232,SMS222:bessa2015probabilistic,SMS225:filipe2015hybrid,SMS235:dowell2016very,SMS239:zhang2016spatial,SMS257:yang2012novel,SMS261:huang2013forecasting}), \textbf{Autoregressive Integrated Moving Average (ARIMA) (3)} (\citeM{SMS176:Oprea2016,SMS252:fernandez2012short,SMS255:pedro2012assessment}), \textbf{Support Vector Regression (SVR) (3)} (\citeM{SMS119:Kramer2013,SMS142:DeLeone2015,SMS150:Heinermann2015}), \textbf{k-Nearest Neighbour (k-NN) (3)} (\citeM{SMS150:Heinermann2015,SMS252:fernandez2012short,SMS255:pedro2012assessment}) 
&
\tiny{
Linear Regression \citeM{SMS109:Hosoda2012119,SMS133:Pravilovic2014}, Multilayer Perceptrons (MLP) \citeM{SMS194:GonzalezOrdiano2017,SMS252:fernandez2012short}, Generalized Autoregressive Conditional Heteroscedasticity GARCH model \citeM{SMS162:Chen2016b,092:chen_wind_2018}, Logistic Regression \citeM{SMS235:dowell2016very},
k-means clustering \citeM{SMS109:Hosoda2012119},  Partial Least Squares Regression (PLSR) \citeM{SMS158:Zhang2015a}, local linear model for time series \citeM{SMS043:Wang2015b}, time-series clustering with Quality Threshold (QT) algorithm \citeM{SMS133:Pravilovic2014}, Support Vector Machine (SVM) \citeM{SMS176:Oprea2016}, Self-Organising Map (SOM) \citeM{SMS119:Kramer2013}, Feedforward Neural Network (FFNN) \citeM{SMS255:pedro2012assessment}, Backpropagation Neural Network (BNN) \citeM{SMS062:Jiang2016b}, Long Short-Term Memory Recurrent Neural Network (LSTM-RNN) \citeM{031:abdel-nasser_accurate_2019}, Entropy-based ANN \citeM{083:ceci_spatial_2019}, Genetic Algorithms \citeM{SMS255:pedro2012assessment}, Random Forests \citeM{SMS150:Heinermann2015}, Ensemble Methods \citeM{SMS109:Hosoda2012119}, Ordinary Least-Squares Fitting \citeM{SMS138:Bessa2015232}, Gaussian Conditional Random Fields (GCRF) \citeM{SMS239:zhang2016spatial}, Particle Swarm Optimization (PSO) \citeM{SMS062:Jiang2016b} }
\\
\hline
C3. Events analysis & 
\textbf{k-means clustering (8)} (\citeM{SMS084:Klinginsmith201775,SMS167:Klinginsmith2016233,SMS155:Marnerides2015,SMS169:Lazzaretti2016a,SMS060:Hosoe2016,006:Nur:2019:CAC:3307772.3331028,051:ISI:000457998600001,052:blasch_dynamic_2018}), \textbf{Hierarchical Clustering Algorithm (HAC) (4)} (\citeM{SMS084:Klinginsmith201775,SMS167:Klinginsmith2016233,SMS126:Fu2014,006:Nur:2019:CAC:3307772.3331028}), \textbf{k-Nearest Neighbour (k-NN) (3)} \citeM{SMS135:Weng201467,049:sial_detecting_2019,066:chong_heuristics-based_2018}, \textbf{DBSCAN (Density Based Spatial Clustering of Applications with Noise) (3)} \citeM{SMS084:Klinginsmith201775,SMS167:Klinginsmith2016233,006:Nur:2019:CAC:3307772.3331028}
& 
\tiny{
Singular Value Decomposition (SVD) \citeM{SMS135:Weng201467,SMS169:Lazzaretti2016a},Support Vector Machines (SVM) \citeM{SMS106:Guarracino2012,SMS084:Klinginsmith201775}, Principal Component Analysis (PCA) \citeM{049:sial_detecting_2019},
Decision Trees \citeM{SMS020:Chen2014a}, x-means clustering \citeM{SMS169:Lazzaretti2016a}, Mean-Shift Clustering (MSC) \citeM{SMS060:Hosoe2016}, d-stream time-series clustering algorithm \citeM{SMS174:Maurya2016a}, time series clustering with Dynamic Time Warping (DTW) \citeM{SMS084:Klinginsmith201775}, Hoeffding Adaptive Tree (HAT) \citeM{SMS223:dahal2015event}, ADaptive sliding WINdow (ADWIN) \citeM{SMS223:dahal2015event}, Piecewise Aggregate Approximation (PAA) \citeM{SMS084:Klinginsmith201775}, model-based anomaly detection \citeM{077:luo_real-time_2018}, LSTM ANN \citeM{051:ISI:000457998600001}, Exponential Smoothing Forecasting Method \citeM{SMS117:Ilic2013a}, Independent Component Analysis \citeM{SMS135:Weng201467}, Kernel Ridge Regression \citeM{SMS135:Weng201467}, Parzen Density Estimator (PDE) \citeM{SMS169:Lazzaretti2016a}, Fuzzy logic \citeM{052:blasch_dynamic_2018}, Monte Carlo simulations \citeM{SMS117:Ilic2013a} }
\\
\hline
C4. Load segregation & 
\textbf{k-Nearest Neighbour (k-NN) (4)} (\citeM{SMS110:Kramer2012,SMS168:Koutitas20161665,030:yang_systematic_2019,072:Laurinec2018210}),
\textbf{Hidden Markov Model (HMM) (3)} (\citeM{SMS140:Bonfigli20151175,SMS212:zeifman2012disaggregation,SMS249:kim2011unsupervised}), \textbf{Support Vector Machines (SVM) (2)} (\citeM{SMS110:Kramer2012,SMS122:Ziekow2013}),  \textbf{Multilayer Perceptrons (MLP)} (2) (\citeM{SMS134:Semwal2014,SMS122:Ziekow2013})
&
\tiny{
Principal Component Analysis (PCA) \citeM{SMS134:Semwal2014,030:yang_systematic_2019}, ARIMA \citeM{072:Laurinec2018210}, Regression Models \citeM{SMS166:Jimenez2016890,072:Laurinec2018210}, Bayes classifier \citeM{SMS134:Semwal2014}, Ensemble Methods \citeM{SMS110:Kramer2012}, Dynamic Time Warping (DTW) \citeM{SMS140:Bonfigli20151175}, Karhunen Loeve (KL) expansion \citeM{SMS144:Dinesh2015}, Ant colony optimization \citeM{SMS146:Gonzalez-Pardo2015}, Optimized Bird Swarm \citeM{038:Wang2019}, ZIP Model-phaselet \citeM{SMS178:Pijnenburg20163408}, Random Forests \citeM{072:Laurinec2018210} }
\\
\hline
C5. Power loads / consumption & 
 \textbf{Support Vector Machines (SVM) (29)} (\citeM{SMS021:Fan20141,SMS026:Yang2014,SMS141:Dagnely2015,SMS202:Moon2017,SMS233:chou2016time,SMS242:fan2017short,SMS066:Liu2016k,SMS087:Li2017a,SMS016:Niska2013,SMS153:Idowu2015554,SMS116:Hou2013170,SMS125:Fattaheian-Dehkordi20141650,SMS151:Hsiao201533,SMS156:Pellegrini2015,SMS163:Dhillon2016a,SMS180:Sreekumar2015,SMS195:He2017254,SMS197:Lei2017,SMS207:Vrablecova2017,SMS259:borges2013evaluating,SMS157:Yu2015a,007:Dong20181,044:ISI:000454335100038,046:Fu201876,047:Wang20196446,056:ISI:000460746500007,065:Yang2018,080:ke_short-term_2019,082:ISI:000425074300010}),
\textbf{Multi Layer Perceptron (MLP) ANN (27)} (\citeM{SMS111:Ramos2012a,SMS008:Lee2012b,SMS021:Fan20141,SMS052:Campos2016,SMS061:Islam2016,SMS094:Usha2017,SMS202:Moon2017,SMS014:Hernandez2013,SMS015:Khan2013b,SMS070:Rana2016118,SMS153:Idowu2015554,SMS107:Hassan2012,SMS120:Marinescu201325,SMS125:Fattaheian-Dehkordi20141650,SMS160:Chemetova2016340,SMS199:Lis2017,SMS204:Rego2017,SMS206:Sun2017,SMS208:Wan2017,SMS211:llanos2012load,SMS157:Yu2015a,SMS231:benedetti2016energy,SMS188:Andersson2017,059:gan_embedding_2018,055:HE2019565,014:kaur_hybrid_2019,058:8652444}), \textbf{Autoregressive Integrated Moving Average (ARIMA) (19)} (\citeM{SMS203:Natividad2017,SMS021:Fan20141,SMS052:Campos2016,SMS098:Yu2017738,SMS233:chou2016time,SMS019:Aparicio2014a,SMS120:Marinescu201325,SMS139:Bianchi20151931,SMS148:Hoeverstad2015,SMS149:Hayes2015a,SMS151:Hsiao201533,SMS185:Alberg2017,SMS193:Gerwig2017}), \textbf{k-means clustering (19)} (\citeM{SMS006:Hernandez2012,SMS203:Natividad2017,SMS003:Kim20111102,SMS103:Kim2011,SMS111:Ramos2012a,SMS187:Alkhatib2017a,SMS016:Niska2013,SMS171:Li2016d,SMS197:Lei2017,SMS206:Sun2017,SMS132:Panapakidis2014137,SMS188:Andersson2017,035:Mohammad20181771,046:Fu201876,048:laurinec_density-based_2019,087:Manojlovic2019140,069:li_long-term_2018,081:dash_short-term_2019,011:ISI:000431531400006,016:ISI:000468028302036,036:Lin2018,043:ISI:000426734600194,044:ISI:000454335100038,046:Fu201876,048:laurinec_density-based_2019,068:Laurinec2019413}), \textbf{Regression Models (16)} (\citeM{SMS090:Paisios2017,SMS021:Fan20141,SMS141:Dagnely2015,SMS186:Alkhatib2017,SMS242:fan2017short,SMS002:Hong2010212,SMS070:Rana2016118,SMS153:Idowu2015554,SMS151:Hsiao201533,SMS160:Chemetova2016340,SMS200:Liu2017730,SMS226:hsu2015identifying,007:Dong20181,047:Wang20196446,067:Song:2018:HDL:3208903.3210283,068:Laurinec2019413,086:gao_probabilistic_2018-1})

&
\tiny{
\textbf{Random Forests (RF) (9)} \citeM{SMS021:Fan20141,SMS186:Alkhatib2017,SMS242:fan2017short,SMS087:Li2017a,SMS193:Gerwig2017,044:ISI:000454335100038,047:Wang20196446,056:ISI:000460746500007,068:Laurinec2019413},
\textbf{Genetic Algorithms (GA) (7)} (\citeM{SMS021:Fan20141,SMS052:Campos2016,SMS061:Islam2016,SMS037:Niska2015,SMS148:Hoeverstad2015,SMS180:Sreekumar2015,SMS190:Bharathi2017}), \textbf{fuzzy c-means clustering (7)} (\citeM{SMS075:Zhang2016f,SMS205:Sharma2017,SMS003:Kim20111102,SMS103:Kim2011,SMS025:Semeraro2014,SMS088:Lu2016b,016:ISI:000468028302036,046:Fu201876}), \textbf{Hierarchical Agglomerative Clustering (HAC) (7)} (\citeM{SMS039:Pal2015,SMS003:Kim20111102,SMS103:Kim2011,SMS111:Ramos2012a,SMS210:Zehetbauer2017,SMS025:Semeraro2014,036:Lin2018}), \textbf{k-Nearest Neighbors (k-NNs) (7)} (\citeM{SMS205:Sharma2017,SMS021:Fan20141,SMS145:Dreiseitl2015,SMS193:Gerwig2017,SMS206:Sun2017,SMS240:zhou2016residential,009:rabie_fog_2019}), \textbf{Decision Trees (7)} \citeM{SMS026:Yang2014,SMS164:Dufour2016a,SMS193:Gerwig2017,SMS197:Lei2017,SMS240:zhou2016residential,056:ISI:000460746500007,065:Yang2018}, \textbf{Fuzzy Logic (6)} (\citeM{SMS041:Son2015,SMS058:Hassan2016,SMS118:Jain2013a,SMS120:Marinescu201325,SMS208:Wan2017,SMS229:shamshirband2015heat}), \textbf{Deep Neural Network (DNN) (6)} \citeM{SMS242:fan2017short,001:nguyen_deepenergy:_2018,005:sun_short-term_2018,007:Dong20181,056:ISI:000460746500007,075:sun_power_2019,080:ke_short-term_2019}, Particle Swarm Optimization (PSO) \citeM{SMS066:Liu2016k,SMS118:Jain2013a,SMS199:Lis2017,014:kaur_hybrid_2019,046:Fu201876}, DBSCAN \citeM{SMS203:Natividad2017,SMS205:Sharma2017,048:laurinec_density-based_2019},  Fuzzy Subtractive Clustering \citeM{SMS017:Pereira2013},  Self-Organizing-Maps(SOM) \citeM{SMS006:Hernandez2012,SMS016:Niska2013,SMS204:Rego2017,SMS211:llanos2012load,SMS132:Panapakidis2014137},  Radial Basis Function (RBF-PCA-WFCM) \citeM{SMS088:Lu2016b}, Wavelet Neural Networks \citeM{SMS120:Marinescu201325,SMS240:zhou2016residential},   Wavelet Transform \citeM{SMS175:Mets2016,SMS070:Rana2016118,SMS148:Hoeverstad2015,011:ISI:000431531400006}, Bayesian Networks \citeM{SMS241:bassamzadeh2017multiscale}, Power Factor Analysis \citeM{SMS090:Paisios2017}, Principal Component Analysis (PCA) \citeM{SMS030:Beaude2015,SMS202:Moon2017,SMS088:Lu2016b,SMS139:Bianchi20151931,068:Laurinec2019413}, Kalman Filter \citeM{SMS086:Lau2017},   g-means clustering \citeM{SMS175:Mets2016},  Simulated Annealing \citeM{SMS061:Islam2016}, MultiVariate Gaussian Distribution Function (MVGDF) \citeM{SMS159:Ali2016}, Montecarlo Simulations \citeM{SMS159:Ali2016,SMS002:Hong2010212}, Pattern Sequence Forecasting (PSF) \citeM{SMS187:Alkhatib2017a}, FA (Factor Analysis) \citeM{SMS202:Moon2017}, Linear Discriminate Analysis (LDA) \citeM{SMS245:li2010classification}, Markov Models \citeM{SMS005:Haider2012,SMS188:Andersson2017,045:Fang2018455}, Bayesian Neural Networks \citeM{029:ISI:000445532600001,043:ISI:000426734600194,065:Yang2018},  Apriori \citeM{SMS066:Liu2016k,SMS197:Lei2017}, Evolutionary Local Kernel Regression \citeM{SMS101:Kramer2010}, GARCH \citeM{SMS116:Hou2013170}, Difference Auto-Regressive (DAR) \citeM{SMS130:Loewenstern2014},  Online Sequential Extreme Learning Machine (OS-ELM) \citeM{SMS171:Li2016d}, Kernel Ridge Regression (KRR) \citeM{SMS193:Gerwig2017,SMS240:zhou2016residential}, Lyapunov optimization \citeM{SMS198:Li2017}, Time Warping Algorithms \citeM{024:Gao2019253}, Ensemble Models~\citeM{SMS021:Fan20141,009:rabie_fog_2019}, Dynamic Mode Decomposition (DMD) \citeM{008:MOHAN2018229}}
\\
\hline
C6. Power quality & 
\textbf{Wavelet Transform (11)} (\citeM{SMS243:ericsti2010wavelet,SMS036:Khan2015a,SMS128:Khan2014a,SMS183:Wang2016,SMS209:Wang20172142,SMS234:cormane2016spectral,SMS251:ning2011wavelet,032:de_andrade_advances_2019,071:huang_missing_2019,090:Bhuiyan2018,073:parvez_online_2018}), \textbf{Principal Component Analysis (PCA) (6)}\citeM{SMS102:Das201195,SMS183:Wang2016,SMS209:Wang20172142,SMS216:mehra2013modes,010:ISI:000445054400035,071:huang_missing_2019}, \textbf{S-Transform algorithm (5)} (\citeM{SMS253:huang2012power,SMS258:biswal2013measurement,SMS260:he2013real,SMS264:biswal2014automatic,040:shamachurn_assessing_2019}), \textbf{Multi-Layered Perceptrons (MLPs) (4)} \citeM{SMS050:Borges2016824,SMS250:lee2011optimal,032:de_andrade_advances_2019,037:ISI:000454332000012}, \textbf{Support Vector Machines (SVM) (4)} \citeM{SMS243:ericsti2010wavelet,SMS209:Wang20172142,040:shamachurn_assessing_2019,073:parvez_online_2018}
&
\tiny{
k-Nearest Neighbour (k-NN) \citeM{SMS250:lee2011optimal,SMS131:Majidpour2014,040:shamachurn_assessing_2019},
k-means clustering \citeM{SMS209:Wang20172142,010:ISI:000445054400035}, Bayesian classifier \citeM{SMS011:Barbosa2013}, Bayesian Network \citeM{SMS046:Ali20161552}, rule-based classification \citeM{SMS256:rodriguez2012rule},  Density Peaks Clustering \citeM{010:ISI:000445054400035}, DBSCAN clustering \citeM{010:ISI:000445054400035}, probabilistic neural network (PNN) \citeM{SMS250:lee2011optimal,SMS253:huang2012power}, decision trees \citeM{SMS050:Borges2016824,SMS260:he2013real}, Deep Neural Network (DNN) \citeM{020:ISI:000458942800091}, Self-Organizing Maps (SOM) \citeM{SMS234:cormane2016spectral}, fuzzy decision tree (FDT)-based classifier \citeM{SMS258:biswal2013measurement}, balanced neural tree \citeM{SMS264:biswal2014automatic} Fuzzy-ARTMAP neural network \citeM{SMS248:decanini2011detection}, Fourier Transform \citeM{SMS260:he2013real,SMS051:Botev2016}, Hilbert transform (HT) \citeM{SMS264:biswal2014automatic}, piecewise compression technique \citeM{SMS035:Eichinger2015a}, nonlinear autoregressive model with exogenous inputs \citeM{SMS067:Liu2016e}, Kalman Filter \citeM{SMS067:Liu2016e}, SZIP algorithm \citeM{SMS218:top2013compressing}, Singular Value Decomposition (SVD) \citeM{SMS183:Wang2016,SMS209:Wang20172142}, Piecewise Aggregate Approximation (PAA) \citeM{SMS209:Wang20172142}, Slack-Referenced Encoding (SRE) \citeM{SMS244:klump2010lossless}, Linear Interpolation Imputation \citeM{SMS177:Peppanen2016}, weighted least square method \citeM{SMS112:Abdel-Majeed2013} }
\\
\hline
C7. Pricing &
\textbf{Support Vector Machine (SVM) (8)} (\citeM{SMS072:Wang2016i,SMS262:shayeghi2013day,SMS263:wu2013new,SMS265:shayeghi2015simultaneous,SMS267:shayeghi2017day,SMS056:Ghasemi201640,018:ghadimi_new_2018,041:barolli_big_2019-1}), \textbf{Wavelet Transform (7)} (\citeM{SMS247:tan2010day,SMS262:shayeghi2013day,SMS265:shayeghi2015simultaneous,SMS267:shayeghi2017day,SMS056:Ghasemi201640,018:ghadimi_new_2018,089:ISI:000426287900006}), \textbf{Multi-Layered Perceptrons (MLPs) (5)} \citeM{SMS266:panapakidis2016day,015:luo_hybrid_2018,018:ghadimi_new_2018,039:ISI:000447404600007,078:ISI:000427121200015}, \textbf{ARIMA (4)} (\citeM{SMS247:tan2010day,SMS262:shayeghi2013day,SMS056:Ghasemi201640,074:Saez-Gallego20185005}), \textbf{Principle Component Analysis (PCA) (4)} \citeM{SMS072:Wang2016i,SMS263:wu2013new,041:barolli_big_2019-1,076:REPL-Li2018-Principalcomponentanalysis}, \textbf{Artificial Bee Colony (ABC) (3)} (\citeM{SMS265:shayeghi2015simultaneous,SMS267:shayeghi2017day,SMS056:Ghasemi201640})
& 
\tiny{
Genetic Algorithms (GA) \citeM{013:rogers_genetic_2019,026:Li2019117,033:Azdemir2018224},
Fuzzy Inference Net (FIN) \citeM{SMS010:Mori2012c}, Fuzzy Self Organising Maps (SOM) \citeM{SMS010:Mori2012c}, fuzzy c-means clustering \citeM{SMS042:Waczowicz2015a}, DBSCAN Clustering \citeM{017:Yang20193374}, Extreme Learning Machine (ELM) \citeM{SMS071:Shrivastava2016},  Deep Neural Network (DNN) \citeM{061:MUJEEB2019101642}, Grey Correlation Analysis (GCA) \citeM{SMS072:Wang2016i,041:barolli_big_2019-1}, Particle Swarm Optimization (PSO) \citeM{089:ISI:000426287900006}, Relevance Vector Machines (RVMs) \citeM{SMS136:Alamaniotis2015}, Linear Regression \citeM{SMS136:Alamaniotis2015}, Autoregressive Moving Average \citeM{SMS136:Alamaniotis2015}, Ensemble Models \citeM{SMS136:Alamaniotis2015,078:ISI:000427121200015}, reference models for price estimation (RMPE) \citeM{SMS238:monteiro2016short}, generalized autoregressive conditional Heteroscedasticity (GARCH) \citeM{SMS247:tan2010day}, data association mining (DAM) algorithms \citeM{SMS254:motamedi2012electricity}, sliding windows algorithms \citeM{054:Almahmoud20193415}, Apriori algorithm \citeM{SMS254:motamedi2012electricity}, Gravitational Search Algorithm (GSA) \citeM{SMS262:shayeghi2013day}, Markov Decision Process \citeM{SMS137:Berlink2015}, Reinforcement Learning \citeM{SMS137:Berlink2015}, Montecarlo simulation \citeM{SMS108:He2012230}, Random Forests (RF) \citeM{041:barolli_big_2019-1}}
\\
\hline
C8. Privacy & \multicolumn{2}{|p{16.0cm}|}{ 
Fuzzy c-means clustering \citeM{SMS215:ford2013clustering}, Random Gaussian Noise \citeM{SMS013:He201367}, Colored Noise \citeM{SMS040:Savi20152409}, Symmetric Geometric Noise \citeM{SMS080:Bao2017}, Secret Sharing Scheme \citeM{SMS081:Guan2017a}, Elliptic Curve Based Data Aggregation (ECBDA) \citeM{SMS095:Vahedi201728}, Collaborative Anonymity Set Formation (CASF) \citeM{SMS184:Afrin2017}, Wavelet-based Multi-resolution Analysis (MRA) \citeM{SMS196:Knirsch2017a}, homomorphic encryption algorithm \citeM{027:ISI:000451814000110}, adversarial strategy algorithm \citeM{SMS219:tudor2013analysis}, fine-grained privacy preservation \citeM{063:Ge2018966}, differentially private aggregated sums \citeM{SMS236:gulisano2016bes}, Integer Linear Optimization (ILP) \citeM{SMS113:Buchmann2013}, Tolerable Deviation algorithm \citeM{SMS115:Ge2013}, partitioning algorithm \citeM{SMS230:tudor2015study}, Haar Wavelet transform \citeM{SMS192:Engel20171710}, Profile Matching protocol using Hamming distance \citeM{SMS181:Unterweger2016}, Kalman filter \citeM{SMS179:Salinas2016}}
\\
\hline
 C9. Security & 
\textbf{Principal Component Analysis (PCA) (8)} (\citeM{SMS012:Esmalifalak2013808,SMS154:Krishna2015,SMS045:Yu20151219,SMS048:Anwar2016180,SMS089:Mohammadpourfard2017242,SMS127:Hao2014,SMS189:Anwar201758,SMS029:Anwar2015a}), \textbf{Support Vector Machines (SVM) (7)} (\citeM{SMS012:Esmalifalak2013808,SMS065:Landford2016,SMS089:Mohammadpourfard2017242,SMS074:Zanetti2016,SMS029:Anwar2015a,003:ganguly_novel_2018,085:Jiao2018})
& 
\tiny{
k-means clustering \citeM{SMS047:An2016240,SMS096:Xu2017a},  Decision Trees \citeM{SMS104:Choi2012,SMS054:Cody20161175},
fuzzy c-means clustering \citeM{SMS089:Mohammadpourfard2017242,042:ISI:000443697900010}, DBSCAN \citeM{SMS154:Krishna2015,SMS089:Mohammadpourfard2017242},  Deep Learning Neural Network \citeM{053:8791598,091:ISI:000429266400030}, Feedforward Neural Network (FFNN) \citeM{SMS063:Kosek2016}, Multi-layer Perceptron (MLP) \citeM{SMS089:Mohammadpourfard2017242}, k-Nearest Neighbor (kNN) \citeM{SMS089:Mohammadpourfard2017242}, clustering-based anomaly detection \citeM{SMS018:Raciti2013}, Ensemble Model \citeM{SMS063:Kosek2016}, Fourier Transform \citeM{050:8646454}, Exponential Smoothing \citeM{SMS078:Andrysiak2017}, Hoeffding Tree \citeM{SMS105:Faisal2012,SMS224:faisal2015data}, Markov Chains \citeM{SMS213:ali2013configuration}, generalized likelihood ratio test (GLRT) \citeM{SMS091:Tang2017172}, Particle Swarm Optimization (PSO) \citeM{SMS096:Xu2017a,021:Chen201973}, Random False Data Attack Detection) \citeM{SMS129:Li2014a}, Weighted Residual Error Method \citeM{SMS152:Hu2015}, Chi-Square Test \citeM{SMS152:Hu2015,SMS189:Anwar201758}, Kullback-Leibler divergence \citeM{SMS191:Bhattacharjee2017}, Generalized Linear Model (GLM) \citeM{SMS191:Bhattacharjee2017}, Cascade Potential Ranking \citeM{SMS201:Mishra2017}, Two Stage Branching Algorithm \citeM{SMS201:Mishra2017}, Colored Petri Nets \citeM{SMS228:liu2015collaborative}, Recursive Least Squares \citeM{SMS057:Han2016a}, Linear Regression \citeM{SMS097:Yip2017230}, Orthogonal Matching Pursuit Cumulative Sum \citeM{070:ISI:000449541500001}, Number Theory Research Unit (NTRU) algorithm \citeM{084:8422731}, Technical Loss Model \citeM{SMS121:Nikovski2013}, Random Matrix Theory \citeM{057:ISI:000424131500014}, Autoregressive Model \citeM{SMS172:Liu2016h}, Locally Regularized Fast Recursive Algorithm \citeM{022:Zhong2018}, Finite Mixture Model \citeM{025:Razavi2019481}, Gradient Boosting \citeM{025:Razavi2019481} }
\\
\hline
C10. SG failures & \multicolumn{2}{|p{16.0cm}|}{
\textbf{k-means clustering (3)} (\citeM{SMS034:DeSantis2015b,SMS220:jiang2014fault,002:sun_distributed_2018}), Basic Sequential Algorithm Scheme (BSAS) \citeM{SMS034:DeSantis2015b}, Genetic Algorithms (GA) \citeM{SMS034:DeSantis2015b,062:deSouzaPereira2018640}, Support Vector Machines (SVM) \citeM{SMS143:DeSantis2015a}, General Regression Neural Networks \citeM{SMS161:Chen20161726}, Deep Neural Network \citeM{004:zitouni_predictive_2019}, Multi-layer Perceptron (MLP) \citeM{023:ISI:000425203500024}, Hidden Markov Model \citeM{SMS220:jiang2014fault}, Ordered Weighted Averaging (OWA) \citeM{SMS085:Kordestani2017}, Radial Basis Functions (RBF) \citeM{SMS085:Kordestani2017}, Logistic Regression \citeM{SMS100:Cai2010642}, Decision Trees \citeM{088:Jana2018387}, k-Nearest Neighbour (k-NN) \citeM{SMS220:jiang2014fault},  Wavelet Transform (WT) \citeM{SMS085:Kordestani2017,088:Jana2018387}, Principal Component Analysis (PCA) \citeM{SMS143:DeSantis2015a}, dynamic optimal synchrophasor measurement devices selection algorithm (OSMDSA) \citeM{SMS165:Jiang20162525}, Wavelet Packet Decomposition \citeM{SMS161:Chen20161726}, Least Square Phasor Estimation \citeM{SMS038:Oubrahim20152735}, Reliability Indexes (SAIDI, SAIFI,...) Thresholds \citeM{SMS064:Kuhi2016}, Multivariate analysis of variance (MANOVA) \citeM{SMS217:nunez2013feature}, fuzzy based Failure Modes and Effect Analysis (FMEA) \citeM{012:Ziga2019507}}
\\
\hline
\end{tabular} 
\end{table*}

\begin{table*}[!htbp]
\renewcommand{\arraystretch}{1.3}
\caption{Main data analysis algorithms applied in the SG domain}
\label{tbl:sg-techniques-examples}
\centering
\begin{tabular}{|p{2.5cm}|p{5.25cm}|p{9.25cm}|}
\hline
\textbf{Technique} & \textbf{Description} & \textbf{Examples Usage in SG} \\
\hline
k-Means clustering (KMC) & An unsupervised clustering technique to cluster items in \textit{k} partitions based on the average distance from cluster centroids. & Used in SG data analysis to cluster user profiles, or energy consumption patterns. For example, Dent \textit{et al.}~\citeM{SMS123:Dent2014} (C1) used SAX (Symbolic Aggregate approXimation) to represent timeseries of power consumption data. Different patterns were clustered with k-means and the clusters quality was then evaluated.\\
\hline
Fuzzy c-means clustering (FCM) & An unsupervised clustering method that allows one item to belong to more clusters. Differs from KMC by considering membership values and a fuzzifier for the level of cluster "fuzziness" & In Benitez \textit{et al.}~\citeM{SMS049:Benitez2016} (C1) power consumption timeseries are clustered using FCM, taking into account the evolution of clusters over time. KMC and FCM with Euclidean or Hausdorff-based similarity measures are found to provide the best results. FCM is also used to locate false data injection attacks in Mohammadpourfard \textit{et al.}~\citeM{SMS089:Mohammadpourfard2017242}~(C9) \\
\hline
Hierarchical clustering (HAC) & An unsupervised clustering method that attempts to build a hierarchy of clusters represented in a dendrogram. & In Chelmis \textit{et al.}~\citeM{SMS032:Chelmis2015}~(C1), HAC is used to determine the customers segmentation for the identification of temporal consumption patterns. In Pal \textit{et al.}~\citeM{SMS039:Pal2015}~(C5), HAC is used for efficient online time series clustering of demand response energy consumption time series data.\\
\hline
Support Vector Machine (SVM) & A supervised learning algorithm for classification and regression. Given a training set, SVM builds a classification/regression model. & Hopf \textit{et al.}~\citeM{SMS059:Hopf2016}~(C1) use SVM to classify households in different profiles according to inputs from smart meter data. Kramer \textit{et al.}~\citeM{SMS110:Kramer2012}~(C2) use SVM in an ensemble (with k-NN) for multi-class classification of household appliances based on power profiles.
Wang \textit{et al.}~\citeM{SMS072:Wang2016i}~(C7) use SVM for energy price classification based on time series multidimensional datasets. \\
\hline
k-Nearest Neighbour (k-NN) & A supervised learning algorithm for classification/regression of new data items based on the k nearest training items by applying a distance function. & Kramer \textit{et al.}~\citeM{SMS110:Kramer2012}~(C2) use k-NN in an ensemble model (with SVM) for non-intrusive load monitoring of appliances. k-NN was found useful in case of large training sets, but more unstable than SVN in the results. In Fan \textit{et al.}~\citeM{SMS021:Fan20141}~(C5) k-NN was used to predict next-day building energy consumption, with the indication that ensemble models provide the best results.
In Jiang \textit{et al.}~\citeM{SMS220:jiang2014fault}~(C10), k-NN is used to support models for SG faults identification.
\\
\hline
Decision Trees (DT) &  A supervised learning algorithm for classification/regression building a decision tree to use as a predictive model.& In Chen \textit{et al.}~\citeM{SMS020:Chen2014a}~(C3), DTs are used for a decision support system based on SG data streams. In Yang \textit{et al.}~\citeM{SMS026:Yang2014}~(C5), DTs are used to predict power consumption in building energy management systems. In Borges \textit{et al.}~\citeM{SMS050:Borges2016824}~(C6), DTs are used to classify power quality disturbances.\\
\hline Multi-Layer Perceptron (MLP) & An artificial neural network used for classification/regression based  conventionally on an input layer, a hidden layer and an output layer. & 
Labeeuw and Deconinck \citeM{SMS007:Labeeuw2012a}~(C5) compared MLP, DT, and SVM for customer sampling based on power consumption patterns. Lee \textit{et al.}~\citeM{SMS008:Lee2012b}~(C5) use MLP to model power consumption forecasting. In Mohammadpourfard \textit{et al.}~\citeM{SMS089:Mohammadpourfard2017242}~(C9), MLP was compared to SVM, k-NN, and DBSCAN algorithms for the detection of SG data injection attacks.\\
\hline
Principal Component Analysis (PCA) & Statistical method to apply data transformation of correlated variables into principal components for data dimensionality reduction. & 
In Chelmis \textit{et al.}~\citeM{SMS032:Chelmis2015}~(C1), PCA is used to reduce multidimensional data to cluster consumption
data in a space of reduced dimension, to simplify both storage and visualization of large amounts of data about power consumption time series.\\
\hline
Wavelet Transform (WT) & A mathematical method to decompose signals into a representation showing trends as a function of time. Divided into Continuous Wavelet Transform (CWT) and Discrete Wavelet Transform (DWT). &
In Rana \textit{et al.}~\citeM{SMS070:Rana2016118}~(C5), WT is used for short-term load forecasting. decomposing electricity load data into components predicted separately. In Khan \textit{et al.}~\citeM{SMS036:Khan2015a}~(C6), WT is used for SG data denoising and compression. In Kordestani \textit{et al.}~\citeM{SMS085:Kordestani2017}~(C10), WT is used for fault location identification from the continuous voltage of the SG buses.
\\
\hline
S-Transform (ST) & A mathematical method for signal decomposition, bringing improvements over the WT. & In Huang \textit{et al.}~\citeM{SMS253:huang2012power}~(C6), ST is used to analyze power quality signals and converted to time-frequency features. In Biswal \textit{et al.}~\citeM{SMS258:biswal2013measurement}~(C6), ST is used to extract features from power quality disturbances.\\
\hline
Autoregressive Integrated Moving Average (ARIMA) & A statistical forecasting technique to predict future values of time series based on past historical trends. & In Oprea \textit{et al.}~\citeM{SMS176:Oprea2016}~(C2), ARIMA is used for power generation forecasting. In Fang \textit{et al.}~\citeM{SMS021:Fan20141}~(C5), ARIMA is used in an ensemble of models to predict next-day building energy consumption.\\
\hline
Hidden Markov Model (HMM) &  A statistical model in which the phenomenon to be modelled is considered a Markov process with externally hidden states. & 
Kim \textit{et al.}~\citeM{SMS249:kim2011unsupervised}~(C4) apply HMM for power load disaggregation, considering the features of appliances use. In Jiang \textit{et al.}~\citeM{SMS220:jiang2014fault}~(C10) HMMs are used for SG faults identification. \\
\hline
\hline
\hline
\end{tabular}
\end{table*}

\textit{C6. Power quality} is focused on several aspects. Power quality disturbances classification deals with the analysis of power disturbance, for example Barbosa \textit{et al.}~\citeM{SMS011:Barbosa2013} provide a classification of power quality disturbances using decision trees. Power data compression is another sub-area: for example, Khan \textit{et al.}~\citeM{SMS036:Khan2015a} provide a data compression and denoising for power signals based on wavelet transforms. Meter placement for quality estimation deals with optimal smart meter location, like in Ali \textit{et al.}~\citeM{SMS046:Ali20161552} that look into optimizing meter placement to identify the best power links for improvement of power quality.
Overall, the C6 category is more focused on the evaluation of data quality by using wavelet transforms techniques, while classification aspects are addressed mainly with the application of neural networks. 

\textit{C7. Pricing} is focused on various aspects of price definition and estimation in the context of SGs. Price forecasting studies are many in this category, like Mori \textit{et al.}~\citeM{SMS010:Mori2012c} that propose a method based on fuzzy logic for electric price zone forecasting. Pricing can also play a role on customer behaviour, like demonstrated by Waczowicz \textit{et al.}~\citeM{SMS042:Waczowicz2015a} that study how dynamic pricing affects households power consumption. The category C7 is more focused on "electricity market" and "pricing forecasts" with classification problems solved by Support Vector Machines (SVM). In this category, smart meters do not play a major role, as the articles are more focused on market characteristics and on the power distribution network.

\textit{C8. Privacy} is dealing with several aspects related to privacy in SG data usage. Privacy preserving data aggregation deals with the preservation of data properties after obfuscation for privacy concerns, like in He \textit{et al.}~\citeM{SMS013:He201367}, that discuss a privacy-preserving approach to protect the privacy of customers maintaining the complete power consumption distribution curve. Data re-identification research looks into how easy is to determine customers or premises starting from anonymized data. Buchmann \textit{et al.}~\citeM{SMS113:Buchmann2013} showed successful re-identification of 68\% of households records from a SG anonimized dataset.

\textit{C9. Security} is focused on security-related aspects such as intrusion detection, false data injection attacks, and energy theft identification. For example, Yu \textit{et al.}~\citeM{SMS045:Yu20151219} studied blind data injection attacks by using Principal Component Analysis (PCA). Kosek and Gehrke \citeM{SMS063:Kosek2016} look into an ensemble-based models for anomaly-based intrusion detection in the SG.  Cody \textit{et al.}~\citeM{SMS054:Cody20161175} propose a Decision Tree approach for fraud detection in consumer power consumption. All the studies provide a view on possible attack vectors, and the countermeasures that can be adopted.

\textit{C10. SG failures} deals with fault status detection, fault type classification, power distribution reliability. For example, De Santis \textit{et al.}~\citeM{SMS034:DeSantis2015b} model the discrimination between fault and normal status of power systems by means of classification and clustering techniques. Kuhi \textit{et al.}~\citeM{SMS064:Kuhi2016} propose a method to measure the power distribution system's reliability from smart meter data. Oubrahim \textit{et al.}~\citeM{SMS038:Oubrahim20152735} look into power system's fault type classification based on PMU estimations.

There is large variability in the aspects covered by the research. Themes that are covered by more articles are power loads forecasting~(89 papers), energy pricing forecasting~(23), forecasting production from renewable power sources~(22), power loads clustering~(21), users power consumption profile clustering~(21), false data injection attacks~(17), users power consumption pattern recognition~(14), power quality disturbances classification~(14), non-intrusive appliance load monitoring~(13), power data compression~(13), energy theft detection~(12), and SG faults detection~(12).

\begin{figure*}[htbp!]
      \renewcommand{\thesubfigure}{C1}
  \subfloat[Customer Profiling]{%
       \includegraphics[width=0.24\linewidth]{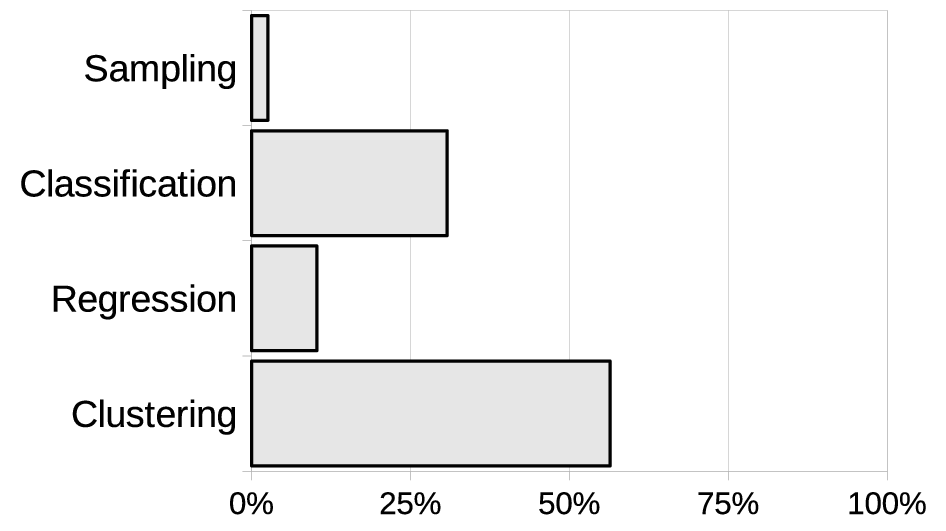}}
      \renewcommand{\thesubfigure}{C2}
  \subfloat[Energy output forecast]{%
        \includegraphics[width=0.24\linewidth]{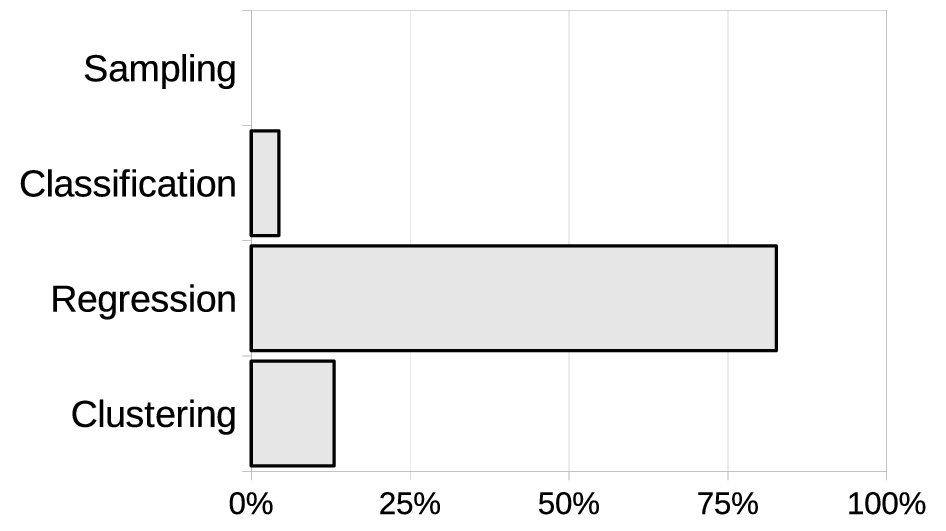}}
     \renewcommand{\thesubfigure}{C3}
  \subfloat[Events Analysis]{%
        \includegraphics[width=0.24\linewidth]{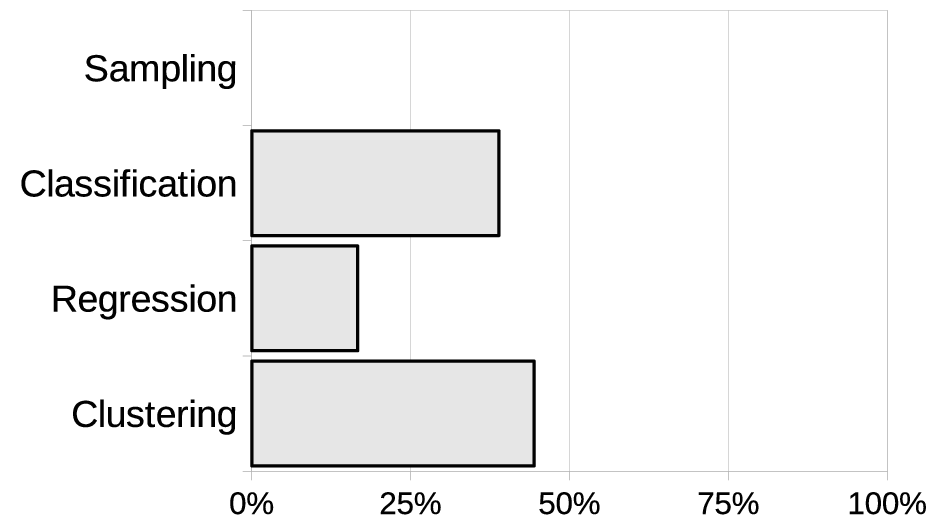}}
     \renewcommand{\thesubfigure}{C4}
  \subfloat[Load Segregation]{%
        \includegraphics[width=0.24\linewidth]{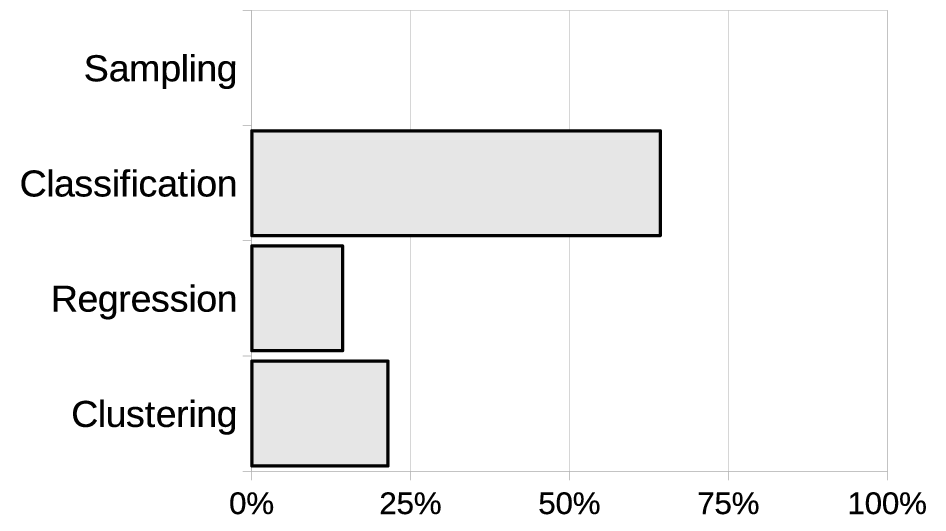}} \\ 
        
  \renewcommand{\thesubfigure}{C5}
  \subfloat[Power load/consumption]{%
       \includegraphics[width=0.24\linewidth]{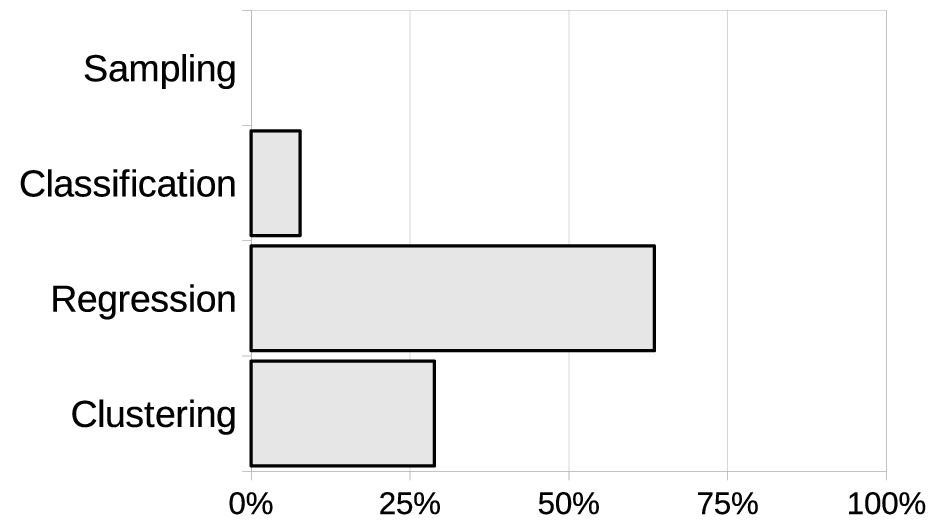}}
  \renewcommand{\thesubfigure}{C6}
  \subfloat[Power quality]{%
        \includegraphics[width=0.24\linewidth]{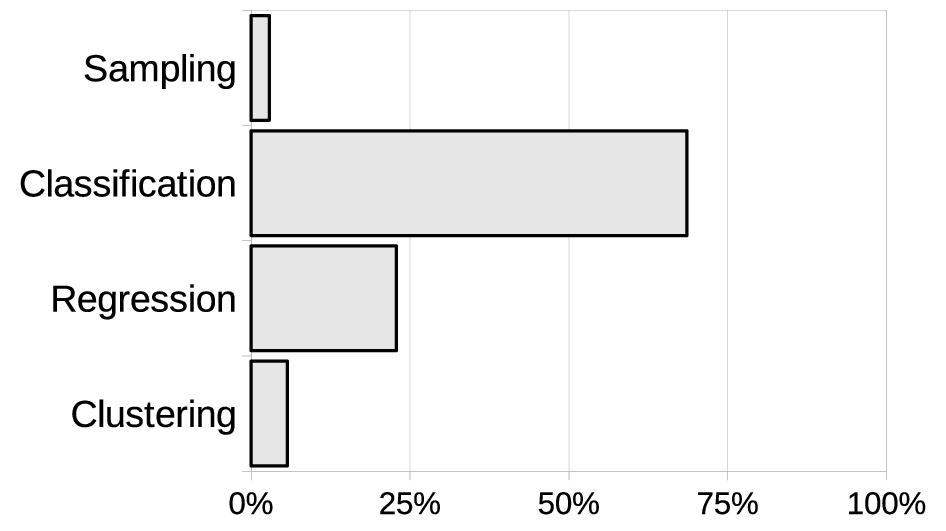}}
  \renewcommand{\thesubfigure}{C7}
  \subfloat[Pricing]{%
        \includegraphics[width=0.24\linewidth]{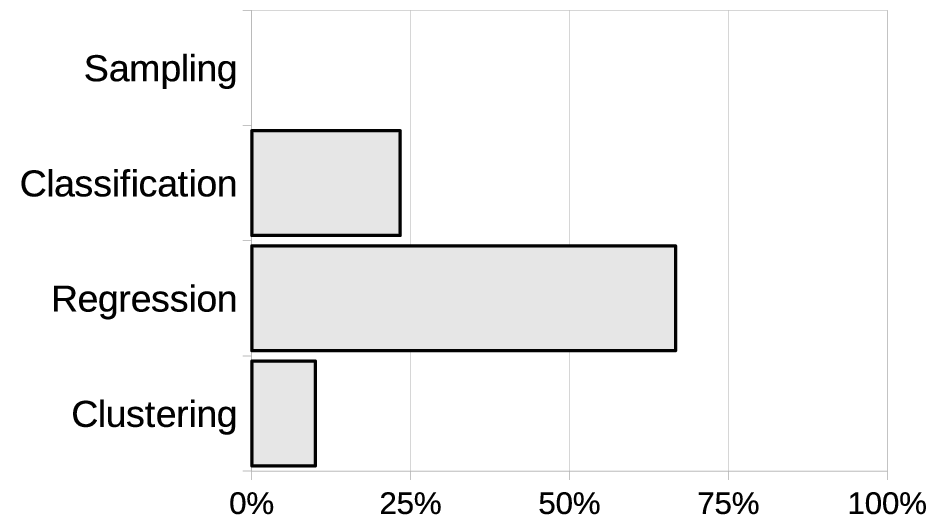}}
  \renewcommand{\thesubfigure}{C8}
  \subfloat[Privacy]{%
        \includegraphics[width=0.24\linewidth]{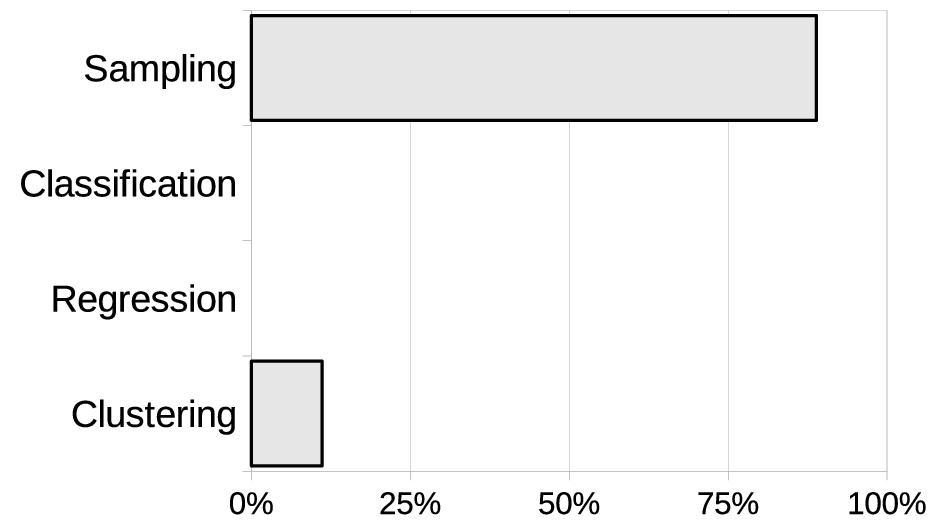}} \\ 
          \renewcommand{\thesubfigure}{C9}
  \subfloat[Security]{%
       \includegraphics[width=0.24\linewidth]{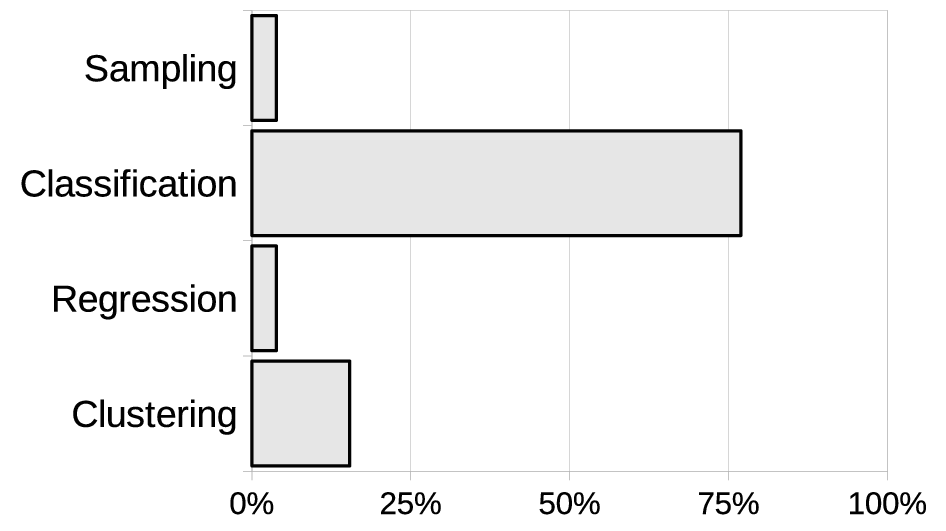}}
  \renewcommand{\thesubfigure}{C10}
  \subfloat[SG failures]{%
        \includegraphics[width=0.24\linewidth]{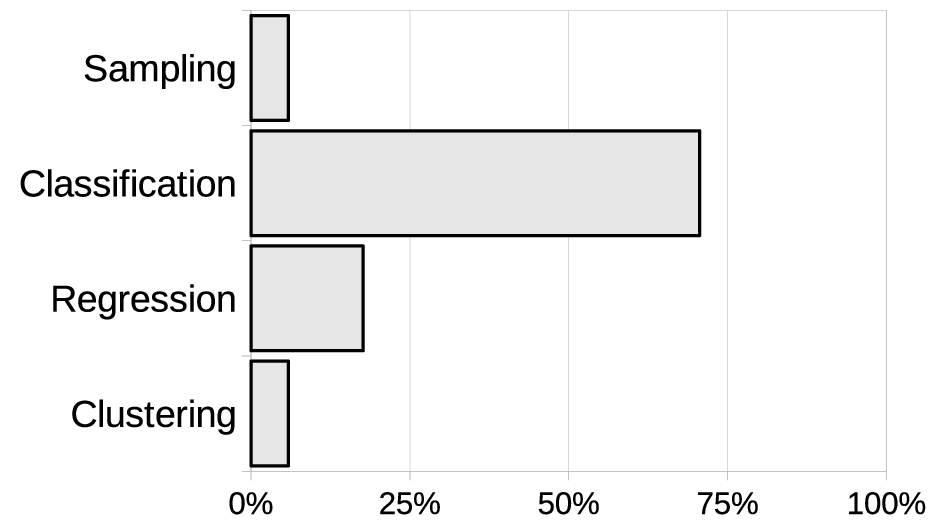}}
        \renewcommand{\thesubfigure}{All}
  \subfloat[Usage of simulations]{%
       \includegraphics[width=0.48\linewidth]{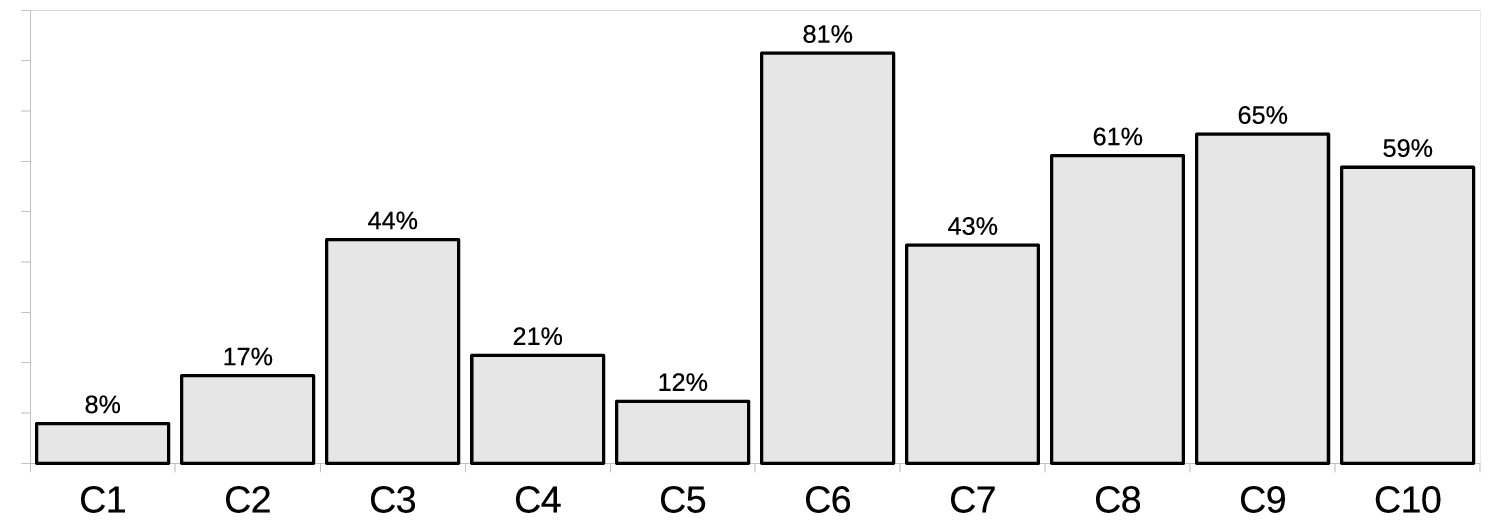}}
  \caption{Research Technique applied (sampling, classification, regression, or cluster analysis) and usage of simulations}
  \label{fig:empirical-technique} 
\end{figure*}

\subsection{RQ2. Which are the reported most used techniques in the identified sub-domains?}
\label{sec:RQ4}
The techniques used in each article and category are mapped in Table \ref{tbl:techniques-mapping}, while a description of the main techniques with some examples of application in SGs are in Table \ref{tbl:sg-techniques-examples}. Overall, there is a huge variety of techniques reported in all the studies.

K-means clustering, together with other clustering techniques such as fuzzy c-means clustering, and hierarchical clustering, are the most applied techniques in the \textit{C1. Customer profiling} category, as most of the studies deal with the definitions of power consumption profiles for customers. Machine learning techniques such as SVM are also widely applied.

Integrated Moving Average (ARIMA) models are the most applied techniques  for \textit{C2. energy output forecasting}, since the articles are mainly focused on solving regression problems.

K-means clustering, and other clustering algorithms are popular in \textit{C3. events analysis} category, as the main focus is on clustering events streams.

Hidden Markov Model (HMM), Support Vector Machines (SVM), k-Nearest Neighbour (k-NN), and Multilayer Perceptrons (MLP) Neural Networks are most used for \textit{C4. load segregation}, mostly for NIALM applications.

Machine learning techniques are mostly used in \textit{C5. power loads/consumption}, with Multi Layer Perceptron (MLP) and Support Vector Machines (SVM) as the most used techniques. There is a variety of other approaches and also linear regression, ARIMA, clustering algorithms are popular.

The \textit{C6. Power quality} category deals with power disturbance classification and algorithms for countermeasures and data compression. Wavelet transform, S-transform, and Principal Component Analysis (PCA) are the most used techniques in the surveyed articles.

In the \textit{C7. Pricing} category, forecasting prices is the main focus: SVM, Wavelet Transforms, ARIMA and Artificial Bee Colony were the most used techniques.

In the \textit{C8. Privacy} category, we did not find single techniques that were the most used. All articles use a variety of approaches, from linear optimization, wavelet transform and linear regression. In this category, due to the focus of research, classification and clustering do not play a major role.

In the \textit{C9. Security} category, Principal Component Analysis (PCA) and Support Vector Machines are the most used techniques. These were applied in particular for several levels of intrusion detection considering SG communication.

In the \textit{C10 SG Failures} category, there is no technique that is most used (apart k-means clustering). There is a variety of techniques, from regression, to PCA, wavelet transforms, SVM, depending on different goal of research that can be prevention or identification of failures at different levels of the SG infrastructure. 

To allow a less fine-grained view, we summarized the techniques used in terms of sampling, classification, clustering, or regression analysis by overall percentage impact in a category (Fig. \ref{fig:empirical-technique}). The categories were defined by looking at articles analyzing part of the data in search for data properties (\textit{sampling}), aggregating data points (\textit{clustering}), solving classification tasks (\textit{classification}), or regression tasks (\textit{regression}). This view highlights areas in which the techniques are more relevant: sampling analysis is more relevant in the \textit{C8. privacy} category, classification is more applied in \textit{C4. load segregation}, \textit{C6. power quality}, \textit{C9. security}, and \textit{C10. SG failures}, clustering is more significant in \textit{C1. customer profiling} and \textit{C3. events analysis} categories, and regression is more targeted to \textit{C2. energy output forecasts}, \textit{C5. power loads/consumption analysis}, and \textit{C7. pricing}.

\subsection{RQ3. Which are the most used software tools / development environments used for data analysis in the identified sub-domains?}

\begin{figure}[!htbp]
\centering

\centering
\includegraphics[width=1.0\linewidth]{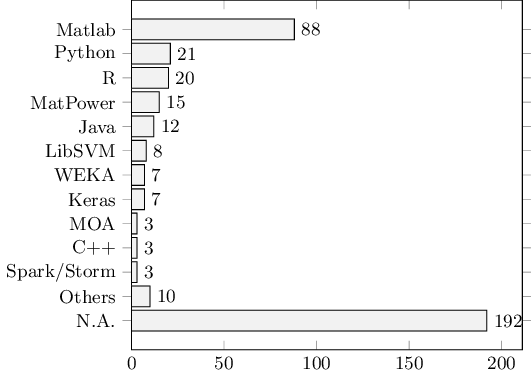}
\caption{Most used tools in the 359 SMS papers (some papers report usage of more software tools / development environments)}
\label{figure:most-used-tools}
\end{figure}

We extracted from the papers the information about the software tools that are reported as more used (Fig. \ref{figure:most-used-tools}). One difficulty in answering this research question is that a large number of articles do not report the software tools used for data analysis ($\sim$53\%, 192/359). The remaining articles ($\sim$47\% 167/359), report the usage of one/more software tool and development environment used to support the data analysis. By far, Matlab (88) is the most reported tool used in the analysis process. Very often, such tool is supported by MatPower for power simulations and LibSVM for machine learning data analysis. As a set of complementary tools, they represent the most widely adopted platform for data analysis in the Smart Grids domain (Fig. \ref{figure:most-used-tools}). 

Python (21) and R (20) represent two development environments widely used for the analysis. In the majority of the cases, scikit-learn is the most reported library used to support machine learning analysis within Python, while for R, the situation is more fragmented, and rarely the packages used are reported. Given the rise of deep learning, there is an increasing usage of Keras (7) with Python. Java (12) was also used in the data analysis, mostly it was used when there was the need to create and support a framework for data analysis that had broader scope than just data processing and algorithms' scripting. Also WEKA (7) was reported for data analysis of SG data: the usage in this case, is mostly about the application of techniques for data mining that are provided by WEKA itself (like  C4.5 (J48) Decision Trees), while other platforms are used when there are needs for more customized (in terms of algorithms) analysis. C++ (3) is limitedly adopted in this context, mostly when some hardware interfacing is needed. When (big)  data streaming aspects are discussed, Apache Storm / Spark (3), Massive Online Analysis (MOA), and Hadoop (in the \textit{"others"} category) were used. Massive Online Analysis (MOA) is an open source framework for large data streams analysis, project that is the complement of WEKA for Big Data analysis.

A methodological aspect worth consideration is the usage of simulations (Fig. \ref{fig:empirical-technique}, \textit{all}). Many papers use data analysis based on numerical simulations, either as complement to datasets analysis or as standalone empirical part of the article. Areas in which simulations are more used are: C6. power quality (81\%), C9. security (65\%), C8. privacy (61\%), C10. SG failures (59\%). In these areas, it can be relatively convenient to generate synthetic datasets based on common properties and evaluate the algorithms. While simulations are less used in other areas: C1. customer profiling (8\%), C5. power loads/consumption (12\%), and C2. energy output forecasts (17\%). The reason is that in these areas there is a large amount of datasets available or that can be gathered from companies participating to common projects, making the adoption of simulations less interesting.

\subsection{RQ4. What is the status of replicability / reproducibility of the studies in terms of datasets used and availability of implemented algorithms?}

In this research question, we investigate the availability of datasets for data analysis and the reproducibility/replicability of the research. The availability of the article alone does not grant the possibility to replicate the whole study. We take two points of views to answer this question: i) one about the availability of the tools / scripts / applications developed for data analysis, and ii) one about the availability of the datasets used (either private or public data repositories).

Availability of algorithms' implementations is quite limited in the 359 papers of the SMS. Overall, only 7/359 ($\sim$2\%) report the availability of the implementations, making difficult for other researchers to reproduce the research. This is a very low number and would require other researchers to re-implement the algorithms/methods reported in the articles, with possibilities of introducing biases in the new implementations.
Articles that report implementations (Table \ref{tbl:papers-with-available-implementation}) are Ceci \textit{et al.} (2019) \citeM{083:ceci_spatial_2019} (C2), Laurinec \textit{et al.} (2019) \citeM{068:Laurinec2019413} (C5), Butunoi \textit{et al.} (2018) \citeM{079:Butunoi20181} (C1), Natividad \textit{et al.} (2017) \citeM{SMS203:Natividad2017} (C5), Bianchi \textit{et al.} (2015)\citeM{SMS139:Bianchi20151931} (C5),  Hoeverstad \textit{et al.} (2015) \citeM{SMS148:Hoeverstad2015} (C5), Bonfigli \textit{et al.} (2015) \citeM{SMS140:Bonfigli20151175} (C4).

\begin{table}[!htbp]
\renewcommand{\arraystretch}{1.3}
\caption{Software implementations source code public repositories}
\label{tbl:papers-with-available-implementation}
\scriptsize
\centering
\begin{tabular}{|p{3.0cm}||p{5.0cm}|}
\hline
\textbf{Article} & \textbf{Public Repository} \\
\hline
Ceci \textit{et al.} (2019) \citeM{083:ceci_spatial_2019} & 
\url{http://doi.org/10.5281/zenodo.1242854}\\
\hline
Laurinec \textit{et al.} (2019) \citeM{068:Laurinec2019413} & 
\url{https://github.com/PetoLau/ClipStream}\\
\hline
Butunoi \textit{et al.} (2018) \citeM{079:Butunoi20181} & 
\url{http://alumni.cs.ucr.edu/~mueen/LogicalShapelet/}\\
\hline
Natividad \textit{et al.} (2017) \citeM{SMS203:Natividad2017} & \url{https://github.com/powertac/powertac-tools}\\
\hline
Bianchi \textit{et al.} (2015)\citeM{SMS139:Bianchi20151931} & \url{https://bitbucket.org/ispamm/distributed-esn}\\
\hline
Hoeverstad \textit{et al.} (2015) \citeM{SMS148:Hoeverstad2015} & 
\url{https://github.com/axeltidemann/load_forecasting}\\
\hline
Bonfigli \textit{et al.} (2015) \citeM{SMS140:Bonfigli20151175} & 
\url{https://nilmtk.github.io}\\
\hline
\hline
\end{tabular}
\end{table}

Ceci \textit{et al.} (2019) \citeM{083:ceci_spatial_2019} deal with renewable energy forecasting and provide the Java implementation of several entropy-based ANNs, together with the datasets used and all the experimental results. Laurinec \textit{et al.} (2019) \citeM{068:Laurinec2019413} focus on the improvement of power consumption forecasting and provide an R-based implementation of a multiple data streams clustering method. Butunoi \textit{et al.} (2018) \citeM{079:Butunoi20181} are interested in the classification of customers power consumption patterns and provide the Java implementation of Logical-Shapelets for time-series classification.
Natividad \textit{et al.} (2017) \citeM{SMS203:Natividad2017}, focus on the usage of machine learning for prediction of energy demand, providing a GitHub repository for data analysis tools for the Power TAC simulation environment.
In Bianchi \textit{et al.} (2015)\citeM{SMS139:Bianchi20151931}, authors report the source code for a recurrent
neural networks \cite{ref:ScardapaneNN2015} used together with PCA decomposition for short-term load forecasting.  Hoeverstad \textit{et al.} (2015) \citeM{SMS148:Hoeverstad2015} provide on GitHub a framework for load forecasting, used in the article about short-term load forecasting with seasonal decomposition.
Bonfigli \textit{et al.} (2015) \citeM{SMS140:Bonfigli20151175} provide a Non-Intrusive Load Monitoring (NILM) Toolkit, a whole framework that can be used by other researchers interested in NIALM (\textit{C4. load segregation}, in our classification). An interesting feature of such toolkit is that it allows to reuse public datasets in the analysis, improving the comparability of research results.

\begin{figure}[H]
\centering
\includegraphics[width=1.0\linewidth]{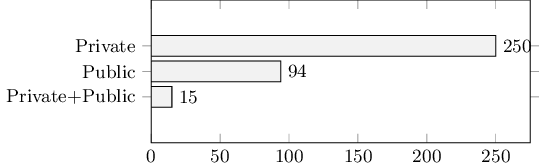}
\caption{Datasets used in the papers (359 papers)}
\label{figure:availability-datasets}
\end{figure}

We looked at the usage of datasets (Fig. \ref{figure:availability-datasets}). In most cases, there are either private, public, or a combination of public and private datasets. In some papers, the used dataset is generated by means of simulations (seen in \textit{RQ3}) or during an experiment. In this latter case, the datasets are always private and not available. 
The majority of the articles  provide the analysis on private datasets (250/359, 70\%), lower number uses only publicly available datasets (94/359, 26\%). Out of the utilized private datasets, 15\% are synthetically generated by means of simulations. There is a small number of articles that utilizes both private and public datasets (15/359, 4\%) by running the same analysis on both to complement the final results.  These numbers do not refer to the absolute number of datasets used in the articles, as articles might use more than one dataset. Furthermore, the public availability of datasets is as reported in the articles---some datasets might not be available anymore (e.g., after some years, the link to the public dataset reported in Semeraro \textit{et al.}~\citeM{SMS025:Semeraro2014} is not reachable anymore. We could not check all, but other similar cases are possible). For a list of publicly available datasets adopted in SG research, see also Chren \textit{et al.}~\cite{ref:chren2018SCSP} and Pereira and Nunes~\cite{ref:pereira2018performance}.

\section{Threats to validity}
 Running an SMS poses several challenges in terms of validity threats~\cite{ref:petersen2015guidelines,ref:budgen2006performing}. 

\textbf{Theoretical Validity.} Theoretical validity refers to threats in building theory out of the observed phenomenon. An SMS cannot cover the whole population of research published in the area, but rather provide a subset \cite{ref:budgen2006performing}. Researcher bias is a theoretical threat in the selection of studies that were part of the SMS. The SMS was conducted by two main researchers, that divided the search process, each one by running the search on a subset of the digital repositories. Results were then merged and controversial papers set apart and reviewed again for inclusion/exclusion. Researchers followed as close as possible the SMS protocols \cite{ref:budgen2006performing}.

Publication bias is another threat: published research might only discuss positive aspects, while negative or controversial aspects/results might not be published. However, this threat is limited, as in an SMS we are more interested in mapping the research than on discussing the findings~\cite{ref:petersen2015guidelines,ref:budgen2006performing}.

\textbf{Descriptive Validity.} Descriptive validity deals with the accuracy of the facts reported by researchers based on the observed phenomenon.
Design of data extraction and data recording can be a descriptive validity threat. Researchers used the JabRef tool (\url{http://www.jabref.org}) to store, manage, and annotate reference lists. Such tool was used for collaboration between the two main researchers performing the SMS, tracking all the selection process.

\textbf{Repeatability / Reproducibility Validity.} Repeatability / Reproducibility Validity deals with how much repeatable (reproducible) the performed process is. Researchers followed the SMS protocol and reporting guidelines \cite{ref:budgen2006performing}, together with sharing all the research artifacts: bibtex files for the steps used in JabRef, and scripts used for the analysis are available at \url{http://dx.doi.org/10.6084/m9.figshare.6804386.v1}. The material also includes additional analyses that could not fit in the paper, like the textual analysis (single terms, bi-grams, and tri-grams) of popular concepts in each category that were used to classify the articles.


\section{Conclusion}

The emergence of the Smart Grid gave rise to many possibilities related to data analysis and to increase the overall intelligence of the infrastructure. Many data analysis approaches were applied in the area to monitor, predict, and provide actionable approaches.

In this paper, we conducted a Systematic Mapping Study (SMS) to get a view of different aspects of data analysis in the context of Smart Grids. Overall, we identified and categorized the articles in ten application sub-domains, focusing on the most common research aspects (e.g., power load forecasting), used techniques (e.g., from time-series to clustering), tool-support status, and reproducibility/replicability levels, research methodology types (e.g., usage of simulations or experiments). We draw conclusions on several aspects:

\noindent (1) From the surveyed articles, we identified ten main sub-domains for SG data analysis: \textit{C1. customer profiling}, \textit{C2. energy output forecasts}, \textit{C3. events analysis}, \textit{C4. load segregation}, \textit{C5. Power loads/consumption analysis}, \textit{C6. power quality}, \textit{C7. Pricing}, \textit{C8. Privacy}, \textit{C9. Security}, \textit{C10. Smart Grid Failures}. In our sample, categories \textit{C1. customer profiling} and \textit{C5. Power loads/consumption analysis} are the most numerous in terms of included articles.

\noindent (2) There is a variety of aspects covered in the different sub-domains. The themes that are covered by more articles are power loads forecasting~(89 papers), energy pricing forecasting~(23), forecasting production from renewable power sources~(22), power loads clustering~(21), users power consumption profile clustering~(21), false data injection attacks~(17), users power consumption pattern recognition~(14), power quality disturbances classification~(14), non-intrusive appliance load monitoring~(13), power data compression~(13), energy theft detection~(12), and SG faults detection~(12).

\noindent (3) There is a large variety of techniques applied for SG data analysis. Each of the SG application sub-domain is focused on different set of research problems and it has impact on the most used techniques. Clustering and classification techniques are very popular in some categories (e.g., \textit{C1. customer profiling}, \textit{C3. events analysis}, \textit{C5. Power loads/consumption analysis}), while wavelet transforms and Principal Component Analysis (PCA) are more popular in other categories (e.g., \textit{C6. power quality}, \textit{C9. security}). The provided table \ref{tbl:techniques-mapping} shows fine-grained details about all the techniques applied in all the surveyed articles.

\noindent (4) Large number of articles do not report any information about the software tools (platforms/programming languages) used for the implementation (192/359, 53\%). Among the articles reporting these details, Matlab (88) (sometimes in combination with MatPower (15) for simulations) is the most used tool. To a lower extent, Python (21), R (20), Java (12), and WEKA (7) are also used. The Keras library (7) in combination with Python seems a popular combination for deep learning articles, as well as libSVM for SVM implementations. Few papers reported about the usage of Big Data frameworks, using mostly MOA (3), and Apache Sparks/Storm (3).

\noindent (5) About replicability/reproducibility of the studies, only 7/359 ($\sim$2\%) articles included in the SMS provide the source code for the implementation of the algorithms that are used for data analysis. This is a very low number to allow other researchers to reproduce the published research. Articles that provide the implementation, usually provide frameworks/toolkits that can be beneficial for other researchers (e.g., Bonfigli \textit{et al.} (2015) \citeM{SMS140:Bonfigli20151175}). Looking at the datasets, 26\% of the articles use only publicly available datasets, giving opportunities of comparability with newly implemented algorithms. There is a small set of articles (4\%) that use both public and private datasets, while the majority of articles is based on private datasets (70\%), making the evaluation of results more complicated for third-party researchers.

Due to the broadness of the area, and the SMS process limitations, we can consider the provided view as a large sample from all existing articles. For each sub-domain, we highlighted the most used techniques, research approaches, and methodologies. The overview can be useful to other researchers performing research in one of the surveyed areas, and for practitioners looking for the applied techniques and methods in the context of SG data analysis studies.



\ifCLASSOPTIONcaptionsoff
  \newpage
\fi



\bibliographystyle{IEEEtranN}

\clearpage

\renewcommand{\IEEEbibitemsep}{0pt plus 2pt}
\makeatletter
\IEEEtriggercmd{\reset@font\normalfont\tiny}
\makeatother
\IEEEtriggeratref{1}

\balance

\renewcommand{\refname}{SMS Articles}

\bibliographystyleM{plain}

\makeatletter
\renewcommand\@bibitem[1]{\item\if@filesw \immediate\write\@auxout
    {\string\bibcite{#1}{M\the\value{\@listctr}}}\fi\ignorespaces}
\def\@biblabel#1{[M#1]}
\makeatother

\tiny{

}


\end{document}